\documentclass[11pt]{article} 
\usepackage{graphicx,floatflt,amssymb,epsf,rotate}  
\usepackage{axodraw4j}
\usepackage{pstricks}
\usepackage{color}  
\def\gtap{\raisebox{-.4ex}{\rlap{$\sim$}} \raisebox{.4ex}{$>$}}

\begin{document}  
\begin{flushright}  
\texttt{hep-ph/yymmnnn}\\  
\texttt{RECAPP-HRI-2013-007}\\  
\end{flushright}  
 
\vskip 30pt

\begin{center}  
{\Large{\bf Boundary Localized Terms in Universal
Extra-Dimensional Models through a Dark Matter perspective}}\\
\vspace*{1cm}  
\renewcommand{\thefootnote}{\fnsymbol{footnote}}  
{ {\sf Anindya Datta${}^{1}$\footnote{email: adphys@caluniv.ac.in}},  
{\sf Ujjal Kumar Dey${}^{1,2}$\footnote{email: ujjaldey@hri.res.in}},  
{\sf Amitava Raychaudhuri${}^{1}$\footnote{email: palitprof@gmail.com}},  
{\sf Avirup Shaw${}^{1}$\footnote{email: avirup.cu@gmail.com}}
} \\  
\vspace{10pt}  
{\small ${}^{1)}$ {\em Department of Physics, University of Calcutta,  
92 Acharya Prafulla Chandra Road, Kolkata 700009, India}}\\ 
  ${}^{2)}$ {\em Harish-Chandra Research Institute, 
Chhatnag Road, Jhunsi, Allahabad  211019, India} \\   
\normalsize

\end{center}

\begin{abstract} 
In universal extra dimension (UED) models with one compactified
extra dimension, a $\mathbf{Z}_2$ symmetry, termed KK-parity,
ensures the stability of the lightest Kaluza-Klein particle (LKP)
which could be a viable dark matter candidate. This symmetry
leads to two fixed points. In non-minimal versions of UED
boundary-localized (kinetic or mass) terms (BLT) for different
fields are included at these fixed points and KK-parity may be
violated.  However, BLTs with same strength at both points induce
a new $\mathbf{Z}_2$ symmetry which restores the stability of the
LKP.  We show that the BLTs serve to relax the bounds set on the
compactification scale in UED by the observed  dark matter relic
density.  At the same time, the precision of the dark matter
measurements severely correlates the BLT parameters of gauge
bosons and fermions.  Depending on the parameter values, the LKP
can be chosen to be the level-1 photon, which is essentially the
$B^{(1)}$, or the level-1 $Z$-boson, basically the $W_3^{(1)}$.
We find that in the latter case the relic density is too small
if the  $W_3^{(1)}$ has a mass $\sim 1$ TeV.  We also explore the
prospects of direct detection of an LKP which matches the
observed dark matter relic density. \\

\vskip 5pt \noindent  
\texttt{PACS Nos:~11.10.Kk, 14.80.Rt, 95.35.+d  } \\  
\texttt{Key Words:~~Universal Extra Dimension, Kaluza-Klein, Dark
Matter}  
\end{abstract}

\renewcommand{\thesection}{\Roman{section}}  
\setcounter{footnote}{0}  
\renewcommand{\thefootnote}{\arabic{footnote}}

\section{Introduction}

With the rapid advancement in observational techniques
cosmological parameters have been measured with unprecedented
precision.  In step, the evidence of a dark matter (DM) component in the
universe has been steadily strengthened. The Planck Satellite
Mission \cite{Planck} reveals that the directly observable baryon
density of the universe is only about $4.9\%$, whereas dark
matter  constitutes about $26.8\%$ of the total energy of the
universe, the rest is due to dark energy. This partitioning is
along the lines indicated earlier by the Wilkinson
Microwave Anisotropy Probe \cite{wmap} and other observations.

Recently, on the basis of measurements made at the International
Space Station, the AMS-02 collaboration \cite{AMS} has published
striking indications for an excess of the positron to electron ratio
which could be an indirect evidence for dark matter. This further
sharpens similar observations from PAMELA \cite{PAMELA} and
Fermi-LAT \cite{Fermi}.

On the particle physics side the standard model (SM) is
a major success in explaining physics up to the TeV scale and has
been repeatedly tested at collider experiments. Very recently the
discovery of a Higgs particle \cite{ATLAS,CMS} has
put the SM on a stronger footing. Yet the model leaves a few
pertinent questions unanswered, e.g., existence of neutrino
masses, matter-antimatter asymmetry, the hierarchy problem, etc.
Moreover, the Standard Model does not solve the DM conundrum as there is no
suitable candidate that can fulfill the requirement.  Although
initially neutrinos or axions were hoped to be the required
DM particle, from present day cosmological observations they
are disfavoured.

The requirement of a dark matter candidate has been one of the
important motivations to go beyond SM. Supersymmetry (SUSY) is by
far the most popular and thus extensively studied
beyond-standard-model scenario. The imposition of a discrete
symmetry -- R-parity -- ensures the stability of the lightest
supersymmetric particle (LSP) which can be a competent DM
candidate. Alternative variants within the SUSY framework predict
different LSPs. For a review of dark matter in the supersymmetric
context see, for example, \cite{jungman}.  A number of DM
candidates have also been put forward in non-supersymmetric
scenarios.  In Little Higgs Models, for example, a conserved
discrete symmetry, known as T-parity, assures the stability of
the lightest T-odd particle, which is typically a heavy photon
\cite{birkedal} and it can be a dark matter candidate.  A common
theme is that in any model attempting to address the dark matter
question, there is some $\mathbf{Z}_2$ symmetry that makes the
lightest symmetry-odd particle stable and, needless to say, that
particle should also satisfy the cosmological observational data,
such as relic density.

Extra-dimensional models, an alternate extension of the SM, also
predict their own  dark matter candidate. In this work we study a
model \cite{acd} where all the SM fields can propagate in the
bulk. We consider models with one extra spacelike flat
compactified dimension, $y$.  If $R$ is the radius of
compactification, this coordinate can be considered to run from 0
to $2 \pi R$. All particles -- scalars, spin-1/2 fermions, and
gauge bosons -- are represented by five-dimensional fields. These
are often conveniently expressed  in terms of towers of
four-dimensional Kaluza-Klein (KK) states.

In the simplest model the zero-modes of the KK-towers are the SM
particles. The KK states of all particles at the
$n$-th level have the same mass, $n/R$.  In addition, a $\mathbf{Z}_2$
symmetry ($y \leftrightarrow -y$) needs to be imposed to ensure
the observed chirality of zero-mode fermions.  The extra
dimension is compactified, in this manner, on an orbifold
$S^1/Z_2$.  The $y \leftrightarrow -y$ symmetry leads to a
conserved KK-parity  $= (-1)^n$, where $n$ is the KK-level.  The
SM particles  ($n$ = 0)  are of even parity
while the KK-states of the first level are odd.  The conservation
of KK-parity ensures that the lightest $n = 1$ particle cannot
decay to SM particles and hence
is a potential dark matter candidate, the Lightest
Kaluza-Klein Particle (LKP). This constitutes what is known as
the Universal Extra Dimension (UED) Model.

The $S^1/Z_2$ orbifold compactification results in fixed points
at $y = 0$ and $y = \pi R$. At these two points one can allow
four-dimensional kinetic and mass terms for the KK-states. In
fact, these terms are also required as counterterms for 
cut-off dependent loop-induced contributions
\cite{georgi} of the five-dimensional theory.  In the minimal
Universal Extra-Dimensional Models (mUED) \cite{cms1, cms2} these
terms are fixed by requiring that the five-dimensional loop
contributions are exactly compensated at the cutoff scale
$\Lambda$ and the boundary values of the corrections, e.g.,
logarithmic mass corrections of KK particles, can be taken to be
zero at the scale $\Lambda$. These  loop contributions can remove
the mass degeneracy among states at the same KK-level $n$.

The viability of the LKP -- a weakly interacting massive particle
(WIMP) -- as a dark matter candidate
satisfying the cosmological relic density constraint has been
examined earlier \cite{mohapatra, taitservant, debasish}. These
studies within the context of UED or mUED at various levels of
sophistication \cite{kongmatchev, burnell} -- e.g., inclusion of
channels of scattering and coannihilation -- indicate that the
$n$ = 1 excitation of the hypercharge gauge boson $B$ could be a
possible candidate for the dark matter of the universe.

In this work we allow  boundary terms to be unrestricted by the
special choice in mUED. In this sense the model can be termed
non-minimal UED (nmUED). However, we do maintain the boundary
terms to be equal at both fixed points.  This will preserve a
discrete $\mathbf{Z}_2$ symmetry which exchanges $y \longleftrightarrow (y
- \pi R)$ and makes the LKP stable.

The compactification radius $R$ and the cut-off $\Lambda$ are the
two basic parameters of these extra-dimensional models and there
have been several explorations of constraints on them in UED  and
its variants. It has been shown that in these models to one-loop
order  electroweak observables receive finite corrections
\cite{db}. Thus it is not unreasonable to compare the predictions
of the theory with experimental data to set bounds on $R$ and
$\Lambda$. For example, from muon $(g-2)$ \cite{nath}, flavour
changing neutral currents \cite{chk,buras,desh}, $Z \to b\bar{b}$
decay \cite{santa}, the $\rho$ parameter
\cite{acd,appel-yee}, and other electroweak precision tests
\cite{ewued, precision}, it is found that $R^{-1}~\gtap~300-600$
GeV.  The possibility of a not too high $R^{-1}$ motivates the
continuing search for signatures of the model at the Tevatron and
the LHC  \cite{collued, LHCout} and also at the ILC or CLIC \cite{ILC} in
the future. 

Here we use the dark matter relic density constraint to place
limits on $R^{-1}$ taking the LKP as the DM candidate. We
consider the impact of the boundary-localized kinetic terms
(BLKT) on the masses of the KK states and, to our knowledge for
the first time\footnote{While this paper was being written up
another paper \cite{newkong} has appeared where the modification
of the couplings due to  five-dimensional wave-functions is 
considered in the context of fermion bulk mass terms along with BLKT.},
also on the KK couplings. In the process the masses of the KK
excitations deviate from their UED value of $n/R$ and the
couplings of these particles exhibit a departure from the
standard model values.  We show that together this makes
additional room for the possible values of the compactification
radius beyond what is permitted in UED.

In the next section, we briefly review the nmUED scenario with
boundary-localized kinetic terms.  To set up the notation this is
followed by a recapitulation of the steps in the estimation of dark
matter relic density.   We come next to  our results where we
identify the regions of parameter space already excluded by the
current limits on the relic density. In the following section we consider the
possibility of direct detection of the nmUED dark matter
candidate for the range of parameters consistent with the relic
density bound. We end with our summary and  conclusions.
Non-trivial nmUED vertices used in this work are collected
together in an Appendix.

\section{Non-minimal UED}
\paragraph*{}

In nmUED one considers kinetic and mass terms localized at the
fixed points. Here  we restrict ourselves to boundary-localized
kinetic terms only \cite{Dvali} - \cite{ddrs}.

Specifically we consider  a five-dimensional theory with additional
kinetic terms localized at the boundaries at $y = 0$ and $y = \pi
R$.  For example, for free fermion fields $\Psi_{L,R}$ whose
zero-modes are the chiral projections of the SM fermions the
five-dimensional action with BLKT is
\cite{schwinn}:
\begin{eqnarray} 
S & = \int d^4x ~dy \left[ \bar{\Psi}_L i \Gamma^M \partial_M \Psi_L 
+ r_f \left\{ \delta(y) + 
\delta(y - \pi R)\right\}  {\phi} ^\dagger _L i \bar \sigma^\mu
\partial_\mu \phi_L
\right. \nonumber \\
&  \left. + \bar {\Psi} _R i \Gamma^M \partial_M \Psi_R
+ r_f \left\{ \delta(y) + 
\delta(y - \pi R) \right\} {\chi} ^\dagger _R i {\sigma}^\mu
\partial_\mu \chi_R
\right]  ,
\label{faction}
\end{eqnarray} 
with $\sigma^\mu \equiv (I, \vec{\sigma})$ and $\bar{\sigma}^\mu
\equiv (I, -\vec{\sigma})$, $\vec{\sigma}$ being the $(2 \times
2)$ Pauli matrices.  Here $r_f$ parametrizes the strength
of the boundary terms which,  for illustrative purposes, we choose
to be the same for $\Psi_L$ and $\Psi_R$ in this section.

It is convenient to express the 
five-dimensional fermion fields using two component chiral
spinors\footnote{The Dirac gamma matrices are in the chiral
representation with $\gamma_5 = diag( -I, I)$.} 
\cite{schwinn}:
\begin{equation} 
\Psi_L(x,y) = \pmatrix{\phi_L(x,y) \cr \chi_L(x,y)} 
=   \sum^{\infty}_{n=0} \pmatrix{\phi_n(x) f_L^{(n)}(y) \cr \chi_n(x) g_L^{(n)}(y)}
\;\; , 
\label{fiveDL}
\end{equation} 
\begin{equation} 
\Psi_R(x,y) = \pmatrix{\phi_R(x,y) \cr \chi_R(x,y)} 
=   \sum^{\infty}_{n=0}  \pmatrix{\phi_n(x) f_R^{(n)}(y) \cr \chi_n(x) g_R^{(n)}(y)} 
\;\;  . 
\label{fiveDR}
\end{equation} 
Below we analyse the case of $\Psi_L$ in detail. The
formulae for $\Psi_R$ will be similar and can be obtained by
making appropriate changes.

Using eq. (\ref{fiveDL}) and varying the action functional  --
eq. (\ref{faction}) -- results in coupled equations for the
$y$-dependent wave-functions, $f_L^{(n)}, g_L^{(n)}$. Following standard
steps\footnote{More details in the same notations and conventions
can be found in \cite{ddrs}.}  one finds:
\begin{equation}
\left[1 + r_f \left\{\delta(y) +  \delta(y - \pi R)\right\}
\right] m_n f_L^{(n)} -
\partial_y g_L^{(n)} = 0,\;\;
m_n g_L^{(n)} + \partial_y f_L^{(n)} = 0, \; (n = 0,1,2, \ldots).
\label{firstordeq}
\end{equation}

Eliminating $g_L^{(n)}$ one obtains the equation:
\begin{equation}
\partial_y^2 f_L^{(n)} + \left[1 + r_f \left\{\delta(y) +  \delta(y
- \pi R) \right \} 
\right] m_n^2 f_L^{(n)} = 0 \, .
\label{fermeq}
\end{equation}
Below we drop the subscript $L$ on the wave-functions and denote
them by $f$ and $g$.

The boundary conditions which we impose are \cite{carena}:
\begin{equation}
f^{(n)}(y)|_{0^-} = f^{(n)}(y)|_{0^+},\;\; f^{(n)}(y)|_{\pi R^+} = f^{(n)}(y)|_{\pi R^-} , 
\end{equation}
\begin{equation}
\frac{df^{(n)}}{dy}\bigg|_{0^+} - \frac{df^{(n)}}{dy}\bigg|_{0^-} = -r_f
m_n^2 f^{(n)}(y)|_{0}, \;\;
\frac{df^{(n)}}{dy}\bigg|_{\pi R^+} - \frac{df^{(n)}}{dy}\bigg|_{\pi R^-} = -r_f
m_n^2 f^{(n)}(y)|_{\pi R} .
\label{fermbc}
\end{equation}
Under the above conditions one gets the solutions:
\begin{eqnarray}
f^{(n)}(y) &=& N_n \left[ \cos (m_n y) - \frac{r_f m_n}{2} \sin (m_n
y) \right] \;,\;\;  0 \leq y < \pi R,   \nonumber \\ 
f^{(n)}(y) &=& N_n \left[ \cos (m_n y) + \frac{r_f m_n}{2} \sin (m_n
y) \right] \;,\;\; -\pi R \leq y < 0.
\label{sol1}
\end{eqnarray}
where the masses $m_n$ for  $n = 0,1, \ldots  $ 
arise from the transcendental equation:
\begin{equation} 
(r^2_f  ~m_n^2 - 4) \tan(m_n \pi R)= 4 r_f  m_n \;.
\label{trans1}
\end{equation}

The solutions
satisfy the {\em orthonormality} relations:
\begin{equation}
\int dy \left[1 + r_f \left\{\delta(y) +  \delta(y - \pi R) \right\}
\right] ~f^{(n)}(y) ~f^{(m)}(y) = \delta^{n m}\;\;.
\end{equation}

The normalisation constant $N_n$  is 
\begin{equation}
 N_n = \sqrt{\frac{2}{\pi R}}\left[ \frac{1}{\sqrt{1 + \frac{r_f^2 m_n^2}{4} 
+ \frac{r_f}{\pi R}}}\right].
\label{norm1}
\end{equation}

Note that the wavefunctions are  combinations of a sine and a
cosine function unlike in the case of mUED where it is one or the
other of these trigonometric functions.  This and the fact
that the KK masses are solutions of  eq. (\ref{trans1}) rather
than simply $n/R$ are at the root of the novelty of nmUED over
the other versions of the theory.  

In our work  we will deal only with the zero-modes and the $n =
1$ KK wave-functions of the five-dimensional fermion fields.

KK wave-functions for scalar and spin-1 fields exhibit very
similar features as above. We illustrate this point
through a five-dimensional gauge field, $G_N ~(N = 0
\ldots 4)$. The action with boundary kinetic terms can be
similarly written as
\begin{equation} 
S = -\frac{1}{4}\int d^4 x \;dy \left[ F_{MN} F^{MN}
+ r_G \left\{\delta(y) 
+ \delta(y - \pi R)\right\} F_{\mu\nu} F^{\mu \nu} \right]  ,
\end{equation} 
where $F_{MN} = (\partial_M G_N - \partial_N G_M)$  and $r_G$ is
the strength of the boundary term (equal at both fixed points)
which is varied in our analysis\footnote{Note that the non-linear term
in $F^{MN}$, which is present if $G_N$ is a non-abelian field,  is
considered a part of the interaction.}. 

The expansion of the gauge  field will be: 
\begin{equation} 
G_{\mu}(x,y)=\sum^{\infty}_{n=0}G_{\mu}^{(n)}(x) a^{(n)}(y),~~~~ 
G_4(x,y) = \sum^{\infty}_{n=0}G_4^{(n)}(x) b^{(n)}(y),  
\end{equation} 
where the functions $a^{(n)}(y)$ and $b^{(n)}(y)$ are determined by the
boundary conditions as discussed below. It is convenient to make
the gauge choice: $G_4 = 0$.

Variation of the action yields the equations satisfied by the
functions $a^{(n)}(y)$:
\begin{equation} 
\partial_y^2 a^{(n)}(y) + \left[1 + r_G \left\{ \delta(y) + \delta(y
- \pi R) \right \} \right]m_n^2 a^{(n)}(y) = 0 .
\label{sol2}
\end{equation} 
We use
the boundary conditions
\begin{equation}
a^{(n)}(y)|_{0^-} = a^{(n)}(y)|_{0^+},\;\; a^{(n)}(y)|_{\pi R^+} = a^{(n)}(y)|_{\pi R^-} , 
\end{equation}
\begin{equation}
\frac{da^{(n)}}{dy}\bigg|_{0^+} - \frac{da^{(n)}}{dy}\bigg|_{0^-} = 
-r_G m_n^2 a^{(n)}(y)|_{0}, \;\;
\frac{da^{(n)}}{dy}\bigg|_{\pi R^+} - \frac{da^{(n)}}{dy}\bigg|_{\pi R^-} = -r_G
m_n^2 a^{(n)}(y)|_{\pi R} .
\label{gaugebc}
\end{equation}

Notice the total similarity between the equation of motion 
and the boundary conditions for a gauge field -- eqns.
(\ref{sol2}) - (\ref{gaugebc}) -- and those
found  earlier for a fermion field as given in eqns.
(\ref{fermeq}) - (\ref{fermbc}). It is but natural that the
wave-functions for the gauge fields can be written down from
those for the fermions by appropriate substitutions.

It is straightforward to verify that the masses $m_n$ in eqn.
(\ref{sol2}) are solutions of:
\begin{equation}
(r^2_G m_{n}^{2}-4)~\tan \left(m_{n}\pi R\right) = 4 r_G m_{n} \; .
\label{trans2}
\end{equation}

Eqn. (\ref{trans1}) satisfied by KK fermions and eqn.
(\ref{trans2}) valid for KK gauge bosons are of identical form.
Though we do not show it here, it applies to the five-dimensional
scalar fields as well. We will next discuss solutions of
transcendental equations of this type. These solutions are the
masses of the KK-excitations if the zero-mode is taken
massless\footnote{As discussed below, the KK modes also receive a
contribution to their masses from spontaneous breaking of the
electroweak symmetry.}, to be compared with $n/R$ for UED. In our
discussion below we will use for the BLKT parameter the general
notation $r$ rather than $r_f, r_G$, etc. It is convenient to
define the dimensionless variables 
\begin{equation}
M_{(n)} \equiv m_n R \;\; {\rm  and} \;\; R_{BLKT} \equiv r/R 
\end{equation}
in terms of which eqs. (\ref{trans1}) and
(\ref{trans2}) can be cast in the form:
\begin{equation}
\left(R_{BLKT}^2M_{(n)}^2 - 4\right) ~\tan (\pi M_{(n)})= 4
~R_{BLKT} M_{(n)}
\end{equation}
which can be readily reexpressed in the factorized form:
\begin{equation}
\left[R_{BLKT}~M_{(n)} + 2 \tan\left(\frac{\pi M_{(n)}}{2}\right)\right]
\left[R_{BLKT}~M_{(n)} - 2 \cot\left(\frac{\pi M_{(n)}}{2}\right)\right] = 0
\label{soln}
\end{equation}
It is easy to convince oneself that the solutions for even $n$
arise from the first factor while those for odd $n$ follow from
the second. The zero mode is the trivial solution from the first
factor and receives no contributions from the BLKT.

\begin{figure}[thb] 
\begin{center} 
\includegraphics[width=4.3cm,height=4.8cm,angle=270]{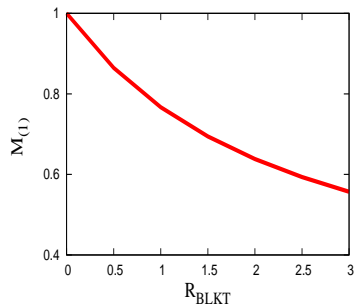}
\caption{Variation of $M_{(1)} \equiv m_1 R$ with the BLKT
strength $R_{BLKT} \equiv r/R$.   The result applies to
fermions, gauge bosons, and Higgs scalars when the BLKTs are
symmetric. Note that larger $R_{BLKT}$ yields a smaller mass. } 
\label{KKmass} 
\end{center} 
\end{figure} 

Here we are interested only in the $n=1$ KK modes.  In Fig.
\ref{KKmass}, we plot $M_{(1)}$ as a function of $R_{BLKT}$ as
obtained from the second factor in eq. (\ref{soln}). For any
chosen $R_{BLKT}$ there is a unique $M_{(1)} = m_1 R$. When we
display the results in the later sections we present  figures
where $R_{B}$ is held fixed and the mass $m_{B^{(1)}}$ is varied by
changing $1/R$. Note that when $R_{BLKT} = 0$, i.e., no BLKT at
all, one gets $m_1 = R^{-1}$, as expected.  Further $m_1$
monotonically decreases as the BLKT strength, $R_{BLKT}$,
increases. Obviously, the LKP will correspond to the $n=1$ KK
excitation of the field with the largest BLKT parameter.

In this work we concentrate on bosonic LKP states; this can be
either $W_3^{(1)}$ or $B^{(1)}$. From our discussion it is clear that
$M_{(1)}$ is determined entirely by the BLKT parameter,
$R_{BLKT}$, and the compactification radius $R$. The gauge
coupling is not involved.  Therefore the earlier discussion
applies for both $W_3^{(1)}$ and $B^{(1)}$ so long as the appropriate
BLKT parameters are used. Which one of these happens to be the
LKP is determined entirely by the choice of the respective BLKT
parameters. In the subsequent sections we investigate the
prospects of these WIMPs playing the role of the dark matter
particle.

We have not included the zero-mode masses in our
consideration till now. We indicate next how they affect the masses
and mixing of $W_3^{(1)}$ and $B^{(1)}$.   The electroweak gauge
boson eigenstates in a five-dimensional theory with BLKT have
been discussed in  the literature \cite{flacke}.  We have
verified that for the range of BLKT parameters which we consider
the states of different KK level, $n$, mix negligibly.  Further,
only if the BLKT parameters for the $B$ and $W$ gauge bosons are
exactly equal (or very nearly so) the mixing between $B^{(1)}$ and
$W_3^{(1)}$ is substantial, it being equal to the zero-mode weak
mixing angle in the case of equality. If $(r_B - r_W)/R$ is of
order 0.1 (or larger) this mixing is negligible.  This can
be verified from the mass matrix which we now discuss.

The mass matrix for the $n=1$ neutral electroweak gauge bosons
including spontaneous breaking of electroweak symmetry as well as
the  extra-dimensional contribution discussed above
is\footnote{Here $m_{W_{3}^{(1)}}$ and $m_{B^{(1)}}$ stand for
the extra-dimensional mass contribution $m_1$, discussed above,
for the $n = 1$  $W_3$ and $B$ states, respectively.}:
\begin{equation}
M_{W_3^{(1)}B^{(1)}} = 
\begin{pmatrix}{
\frac{g_2^2v_0^{2}}{4} \frac{S_{W}}{S_H} 
I_{W_{3}W_{3}}+{{m^2_{W^{(1)}_{3}}}}
&
-\frac{g_{2}g_{1}v_0^{2}}{4} \frac{\sqrt{S_{W}S_{B}}}{S_H}
I_{W_{3}B}\cr
& \cr
-\frac{g_{2}g_{1}v_0^{2}}{4} \frac{\sqrt{S_{W}S_{B}}}{S_H}
I_{W_{3}B}&
\frac{g_1^2v_0^{2}}{4} \frac{S_{B}}{S_H}
I_{BB}+{{m^2_{B^{(1)}}}}
}\end{pmatrix}, \nonumber
\label{mmatrix}
\end{equation}
where
\begin{equation} 
I_{ij}=\int^{\pi R}_{0} 
\left(1+r_{h}\{\delta(y)+\delta(y-\pi R)\}\right)
a^{(1)}_{i}(y) a^{(1)}_{j}(y)  dy, \;\; (i, j = W_{3},B)  .
\end{equation}
Above, $r_{h}$ is the strength of the  Higgs
scalar BLKT and $R_h = r_h/R$. Also,
\begin{equation} 
a^{(1)}_{W_{3}}(y) =
N_{W_{3}}^{(1)}\left[\cos\left(m_{W_{3}^{(1)}}\ y\right) -
\frac{r_{W} m_{W_3^{(1)}}}{2}\sin\left(m_{W_{3}^{(1)}}\ y\right)\right], 
\end{equation}
and
\begin{equation} 
a^{(1)}_{B} (y) = N_{B}^{(1)}\left[\cos\left(m_{B^{(1)}}\ y\right) -
\frac{r_{B}m_{B^{(1)}}}{2}\sin\left(m_{B^{(1)}}\ y\right)\right] , 
\end{equation}
with $N_{W_{3}}^{(1)}$, $N_{B}^{(1)}$ being  normalisation
factors as in eqn. (\ref{norm1}).

The five-dimensional gauge couplings $\hat{g_{2}}, \hat{g_{1}}$ and the
vacuum expectation value (vev) $v_5$ are related to the four-dimensional
couplings $g_{2}, g_{1}$ respectively and the vev $v_0$  through 
\begin{equation}
\hat{g_{2}} = g_{2}\sqrt{\pi R ~S_W}  \;, \; ~\hat{g_{1}} = g_{1}\sqrt{\pi R ~S_B} \; , \; v_5 =
v_0/\sqrt{\pi R ~S_H} \;,  
\end{equation}
where

\begin{equation}
S_{W} = \left(1+\frac{R_W}{\pi}\right), \;\; 
S_B = \left(1+\frac{R_B}{\pi}\right),\;\;
S_H = \left(1+\frac{R_h}{\pi}\right).
\end{equation}

A few comments about the mass matrix in eqn. (\ref{mmatrix}) are
in order. The matrix is given in the $W_3^{(1)} - B^{(1)}$
basis\footnote{We have checked that mixing with states of $n \neq
1$ is very small.}.  To estimate the relative magnitudes of the
terms  notice that the $S_i$ are ${\cal O}(1)$ as
are the overlap integrals $I_{ij}$. Therefore the contributions
to the mass matrix from the symmetry breaking are ${\cal
O}(v_0^2)$. The extra-dimensional contributions, ${m^2_{G^{(1)}}}$,
are of the order of $(1/R)^2$ and are always dominant by far.  As
a consequence these terms determine the mass eigenvalues and the
mixing  is negligible for $(R_W - R_B) \sim 0.1$ or
larger\footnote{If $R_W = R_B$ then the dominant diagonal terms
are equal and do not affect the mixing and simply make a constant
shift of the masses of the eigenstates. In this case the mixing
between $W_3^{(1)}$ and $B^{(1)}$ is just as in the Standard Model with
$\tan \theta = g'/g$.}.  So, in our discussion below we take
$B^{(1)}$ and $W_3^{(1)}$ to be the neutral electroweak gauge
eigenstates.  The contributions to the masses themselves from
electroweak symmetry breaking are insignificant and hence
dropped. We have verified that for the cases of our interest
there is no significant dependence of the results on the Higgs
scalar BLKT strength, $r_h$. In our calculations we keep $R_h =
0.1$ throughout.

\section{Standard calculation of relic density}
\paragraph*{}

In this section we briefly summarize the standard calculation of
relic density. This will facilitate us in defining the notations as
well as to put the calculation in the nmUED model in a proper
context.  For a more detailed elaboration of this formulation 
the reader is  referred to \cite{taitservant, kongmatchev}.  

Let us consider the spectrum of the $n=1$ KK-level of nmUED in which
the particles are specified by $\alpha_i, (i = 0, 1, 2, \ldots)$
ordered sequentially in mass. $\alpha_0$ is identified with
$B^{(1)}$, which is the lightest\footnote{In nmUED, one has the
flexibility to arrange $W_3 ^{(1)}$ or $\nu^{(1)}$ or $H^{(1)}$
to be the LKP. In this work we consider only the $W_3^{(1)}$ or
$B^{(1)}$ as the dark matter candidate.  In this
section we stick to the $B^{(1)}$ LKP case.} of these states and
hence stable.  It is also  massive and weakly interacting.
Consequently it is a natural candidate for being the dark matter
WIMP.  Apart from $\alpha_0$ all other particles are unstable and
ultimately decay to $B^{(1)}$ in association with SM particles.
Thus from any number $N_i$ of $\alpha_i$ particles we end up
getting exactly $N_i$ $\alpha_0$ at the end of the decay chain.
Consequently, to determine the relic density of $\alpha_0$ it
would be meaningful to study the evolution of $n (\equiv \sum_i
n_i)$ instead of each number density $n_i$ separately.

The evolution (with time or temperature) of the number density $n$ is
governed by the Boltzmann transport equation:
\begin{equation}
\frac{d n}{ d t} = -3 Hn - \langle \sigma_{eff} v \rangle ( n^2 - n^2_{eq})\ .
\label{boltzman}
\end{equation}
here $H$ is the Hubble parameter  and $n^{eq}$ the number
density at thermal equilibrium. $ \langle \sigma_{eff} v \rangle$
is the thermally averaged relative velocity times the effective
interaction cross section of $\alpha_0$ with other particles in
the spectra.

More precisely, $\sigma_{eff}$ can be defined as, 
\begin{equation}
\sigma_{eff}(x) = \frac{1}{g_{eff}^2} \sum_{ij}^N \sigma_{ij} F_i F_j 
\ ,
\end{equation}
with
\begin{equation}
F_i(x)   = g_i (1+\Delta_i)^{3/2} \exp(-x \Delta_i)\,\,  {\rm
and} \,\, g_{eff}(x)   = \sum_{i=1}^N F_i\ . 
\end{equation}
Here
\begin{equation}
\Delta_i     = \frac{m_i - m_0}{m_0}\,\,, \,\,x =
\frac{m_0}{T}\, ,
\end{equation}
with $\sigma_{ij}\equiv \sigma(\alpha_i \alpha_j\to SM)$ and
$g_i$ is the number of degrees of freedom for the particle
$\alpha_i$ taking part in the annihilation or coannihilation
process.

\setcounter{table}{0}  
\renewcommand{\thetable}{\arabic{table}}  

\begin{table}
\vskip -20pt
\begin{center}
\begin{tabular}{|l|l|}
\hline
 \textbf{$B^{(1)}$ annihilation} & 
 \textbf{$B^{(1)}$-lepton scattering} \\ \hline \hline
$ B^{(1)} B^{(1)} \longrightarrow f \bar{f}$ & $\nu_L^{(1)} B^{(1)} \longrightarrow W^{+} l^{-}$ \\ \hline
 $ B^{(1)} B^{(1)} \longrightarrow h^{+} h^{-}$ & $\bar{\nu}_L^{(1)} B^{(1)} \longrightarrow W^{-} l^{+}$\\ \hline
 & $\nu_L^{(1)} B^{(1)} \longrightarrow Z \nu_{l}$ \\ \hline
 & $\bar{\nu}_L^{(1)} B^{(1)} \longrightarrow Z \bar{\nu}_{l}$\\ \hline
 & $l_L^{(1)-} B^{(1)} \longrightarrow W^{-} \nu_{l}$ \\ \hline
 & $l_L^{(1)+} B^{(1)} \longrightarrow W^{+} \bar{\nu}_{l}$ \\
\hline
 &  $l_L^{(1)-}B^{(1)}\longrightarrow Z l^{-}$ \\
\hline
 &  $l_L^{(1)+}B^{(1)} \longrightarrow Z l^{+}$ \\
\hline
\end{tabular}
\caption{The $B^{(1)}$ annihilation and $B^{(1)}$-lepton
scattering processes which contribute to the relic density
calculation. The cross sections involve, besides the SM
couplings, the vertices in secs.
\ref{b1h1h0}, \ref{b1b1h0h0}, and \ref{b1f1f0}.}
\label{t:1}
\end{center}
\end{table}

The right-hand-side of eqn. (\ref{boltzman}) has two terms. The
first  (proportional to $H$) accounts for the reduction of the
number density due to the expansion of the universe, and the second
one for the decrease (increase) of $n_i$ due to its interaction
with the other particles present in the spectrum.  

It has been noted  \cite{taitservant, kongmatchev} that in
case all the $\Delta_i$ are larger than 10\% $\sigma_{eff}$ can
be well estimated by calculating only the annihilation of
$\alpha_0$ with its own anti-particle to SM particles.  The
co-annihilation (processes involving  other KK-particles which 
cascade decay ultimately to the LKP)
contribution is important only when one or more $\Delta_i$ are
less than 10\%. 

In the non-relativistic limit (which is appropriate in our
calculation as we are interested to calculate the number density
in an epoch when average temperature is less than
the mass of the WIMP) one can use the approximation $\langle
\sigma_{eff} v \rangle \sim a_{eff}(x) +
b_{eff}(x)\, v^2 + {\cal O}(v^4)$. 

After calculating the relevant $\sigma_{ij}$ one can numerically
integrate eqn. (\ref{boltzman}) to obtain the number density of
the DM particle at the present epoch.  Instead, here we use an
approximate formula \cite{taitservant} for the number density 
(normalised to the critical density at the present epoch) given
by:

\begin{equation}
\Omega_\alpha h^2 \approx \frac{1.04 \times 10^9/1 {\rm GeV}}{M_{Pl}}
\frac{x_F}{\sqrt{g_\ast(x_F)}} ~\frac{1}{I_a+3 I_b/x_F }\ ,
\label{approxomega}
\end{equation}
where $M_{Pl}$ is the Planck mass and
\begin{eqnarray}
I_a &=& x_F \int_{x_F}^\infty a_{eff}(x) x^{-2} d x \ ,
\\
I_b &=& 2 x_F^2 \int_{x_F}^\infty b_{eff}(x) x^{-3} d x\ .
\label{iaib}
\end{eqnarray}
Here $x_F$ is  the  freeze-out temperature and can be solved from
the following equation,
\begin{equation}
x_F = \ln \left (\frac{15}{8} \sqrt{\frac{5}{2}}
~\frac{g_{eff}(x_F)}{2\pi^3} ~\frac{m_0 M_{Pl}
[a_{eff}(x_F)+6b_{eff}(x_F)x_F^{-1}]} {\sqrt{g_\ast(x_F) x_F}}
\right )\ .
\end{equation} 
Above, $g_{\ast} (x_F)$ accounts for the relativistic degrees of
freedom at the freeze-out temperature for which we have used the value
of 92. For the parameter ranges used in this
analysis $x_F$ is found to be around 23 - 26.


\begin{table}
\vskip -20pt
\begin{center}
\begin{tabular}{|l|l|}
\hline
 \textbf{$\nu^1$ annihilation and scattering} & 
 \textbf{$l^1$ annihilation and scattering} \\ \hline \hline
$\nu_l^{(1)}\bar{\nu}_l^{(1)} \longrightarrow f \bar{f}$ & 
$l_L^{(1)+}l_L^{(1)-} \longrightarrow h^{+} h^{-}$ \\ \hline
$\nu_l^{(1)}\bar{\nu}_l^{(1)} \longrightarrow h^{+} h^{-}$ &
$l_L^{(1)+}l_L^{(1)-} \longrightarrow ZZ+Z\gamma + \gamma \gamma$ \\ \hline
$\nu_l^{(1)}\bar{\nu}_l^{(1)} \longrightarrow ZZ$ & 
$l_L^{(1)+}l_L^{(1)-} \longrightarrow W^{+}W^{-}$ \\ \hline 
$\nu_l^{(1)}\bar{\nu}_l^{(1)} \longrightarrow W^{+}W^{-}$ & 
$l_L^{(1)\pm}l_L^{(1)\pm} \longrightarrow l^{\pm}l^{\pm}$ \\ \hline
$\nu_l^{(1)}\nu_l^{(1)} \longrightarrow \nu_l \nu_l$ & 
$l_L^{(1)\pm}l_L^{(1)\prime\pm} \longrightarrow l^{\pm}l^{\prime \pm}$ \\ \hline 
$\nu_l^{(1)}\nu_{l^{\prime}}^{(1)} \longrightarrow \nu_l \nu_{l^{\prime}}$ & 
$l_L^{(1)\pm}l_L^{(1)\prime\mp} \longrightarrow l^{\pm}l^{\prime \mp}$ \\ \hline
$\nu_l^{(1)}\bar{\nu}_{l^{\prime}}^{(1)} \longrightarrow \nu_l \bar{\nu}_{l^{\prime}}$ &
$l_L^{(1)+}l_L^{(1)-} \longrightarrow f \bar{f} \; \; or \; \; l_{R}^{+}l_{R}^{-}$ \\ \hline
$\nu_l^{(1)}\bar{\nu}_{l^{\prime}}^{(1)} \longrightarrow l^{-}l^{\prime +}$ &
$l_L^{(1)+}l_L^{(1)-} \longrightarrow \nu_{l}\bar{\nu}_{l} \; \; or \; \; l_{L}^{+}l_{L}^{-}$
\\ \hline 
$\nu_l^{(1)}\bar{\nu}_l^{(1)} \longrightarrow l^{+}l^{-}$ &
$l_L^{(1)}\bar{l}_L^{(1)\prime} \longrightarrow \nu_{l}\bar{\nu}_{l^{\prime}}$ \\ \hline
\end{tabular}
\caption{The $\nu^{(1)}$ and $l^{(1)}$ annihilation and
scattering processes which contribute to the relic density
calculation. The cross sections involve, besides the SM
couplings, the vertices in secs. \ref{b1f1f0} and \ref{w1f1f0}.}
\label{t:2}
\end{center}
\end{table}

At this point, given a model one is well-equipped to calculate the
relic particle density.  In UED (also in nmUED) there are two main
classes of reactions which contribute to $\sigma_{eff}$; namely, the
annihilation of relic particles into the SM particles and the
co-annihilation processes (explained above).  It is to be noted
that in UED the mass spectrum depends only on one parameter
$R^{-1}$.  In contrast, in nmUED masses of the KK-excitations
have strong dependence on all the BLKT parameters.  Thus it is  the
choice of the BLKT parameters which determines whether or not the
mass of any KK-excitation would lie close to the
relic particle mass ($\Delta_i \leq 10$\%) so that their
co-annihilation process could contribute significantly. In our
analysis we have chosen the BLKT parameters for quarks, gluons
and right-handed leptons such that their masses are sufficiently
larger than the $B^{(1)}$ mass and their contributions to
co-annihilation are negligible.  For this choice only the
coannihilation of KK-excitations of the left-handed lepton
doublets are relevant.

In Tables \ref{t:1} - \ref{t:3} we have listed the reactions
which are used to calculate $\sigma_{eff}$.  Some of the
couplings among the $n=1$ KK-modes and SM particles get modified
in nmUED due to the non-trivial wave-functions of the
KK-excitations in the fifth-dimension. (We have listed such
interactions with the Feynman rules in Appendix A.)  Consequently,
the cross sections of some of the processes contributing in
$\sigma_{eff}$ will be accordingly scaled up or down with respect
to their UED values.


\begin{table}
\vskip -10pt
\begin{center}
\begin{tabular}{|l|l|}
\hline
\multicolumn{2}{|c|}{\textbf{$\nu^1$-$l^1$ scattering}} \\
 \hline \hline
$\nu_l^{(1)}l_L^{(1)\prime} \longrightarrow \nu_{l}l^{\prime}$ &
$l_L^{(1)-}\bar{\nu}_L^{(1)} \longrightarrow f \bar{f}^{\prime}$ \\ \hline
$l_L^{(1)-}\bar{\nu}_L^{(1)} \longrightarrow h^{-} h^{0}$ & 
$l_L^{(1)-}\bar{\nu}_L^{(1)} \longrightarrow \gamma W^{-} $ \\ \hline
$l_L^{(1)-}\bar{\nu}_L^{(1)} \longrightarrow Z W^{-} $ & 
 $l_L^{(1)-}\bar{\nu}_L^{(1)} \longrightarrow l^{-}\bar{\nu}_{l}$ \\ \hline
$l_L^{(1)-}\nu_L^{(1)} \longrightarrow l^{-}\nu_{l}$ &
$\nu_L^{(1)}l_L^{(1)\prime} \longrightarrow \nu_{l^{\prime}}l^{-}$ \\ \hline
$\nu_L^{(1)}l_L^{(1)\prime} \longrightarrow \nu_{l}l^{\prime -}$ &
$\bar{\nu}_L^{(1)}l_L^{(1)\prime} \longrightarrow \bar{\nu}_{l}l^{\prime -}$ \\ 
\hline
\end{tabular}
\caption{The $\nu^{(1)}-l^{(1)}$ scattering processes which
contribute to the relic density calculation. The cross sections 
involve, besides the SM couplings, the vertices in secs. \ref{b1f1f0}
and \ref{w1f1f0}.}
\label{t:3}
\end{center}
\end{table}

\section{Dark Matter relic density in nmUED} 

We are now
ready to present the results and compare it with the current
observed value\footnote{We use of $\Omega h^2 = 0.1198
\pm 0.0026$, the $ 1\sigma$ results from Planck.} of $\Omega h^2$
from Planck data \cite{Planck}.  A similar exercise in the
framework of  UED (when $B^{(1)}$ is the candidate for dark
matter) yields a narrow range\footnote{It has been pointed out
\cite{kaki} that if the contributions from $n$=2 excitations are
included then this limit is pushed up above 1 TeV.} \cite{kongmatchev}  of allowed
values -- 500 - 600 GeV -- for the compactification scale
$R^{-1}$.  However we will see that in nmUED this bound on
$R^{-1}$ will be relaxed significantly\footnote{We also make remarks
in passing about the constraint set on the LKP mass or equivalently
the compactification scale by the limit from overclosure of the
universe.}. We take up the cases of $B^{(1)}$
and $W_3^{(1)}$ LKP separately.

\subsection{Results for $B^{(1)}$ LKP} 

\begin{figure}[thb] 
\begin{center} 
{\includegraphics[width=4.8cm,height=4.8cm,angle=270]{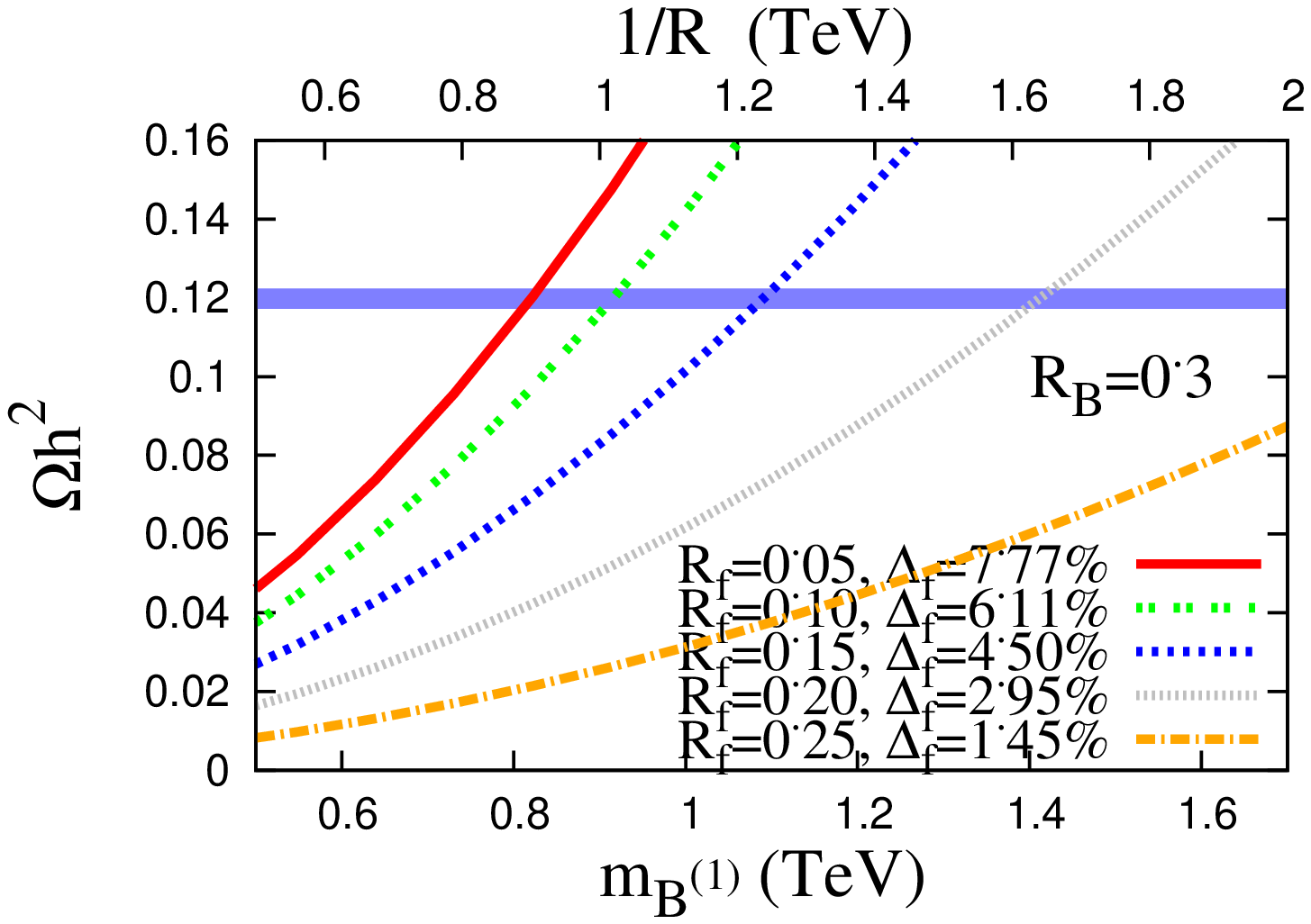}
\includegraphics[width=4.8cm,height=4.8cm, angle=270]{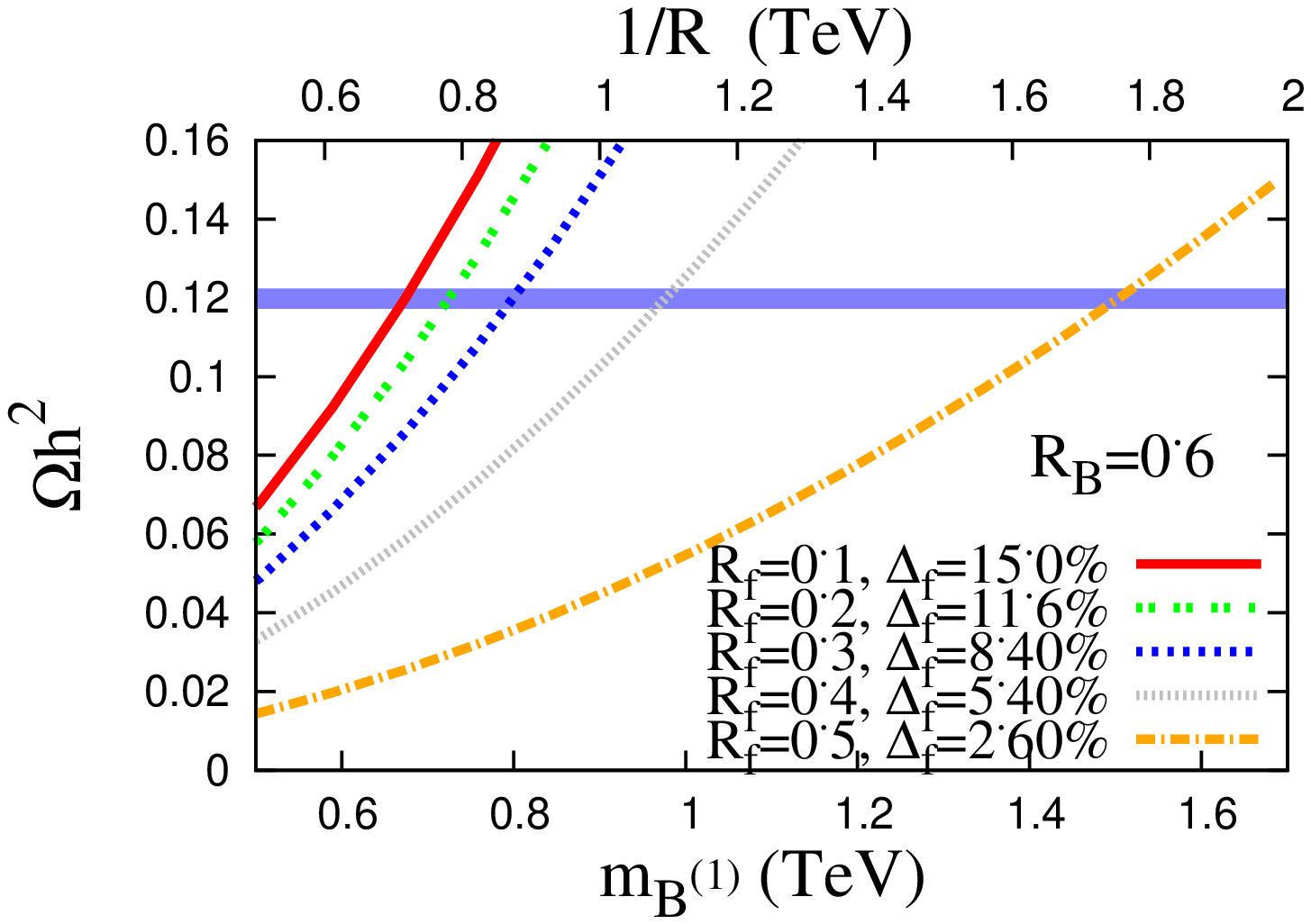}}
\includegraphics[width=4.8cm,height=4.8cm, angle=270]{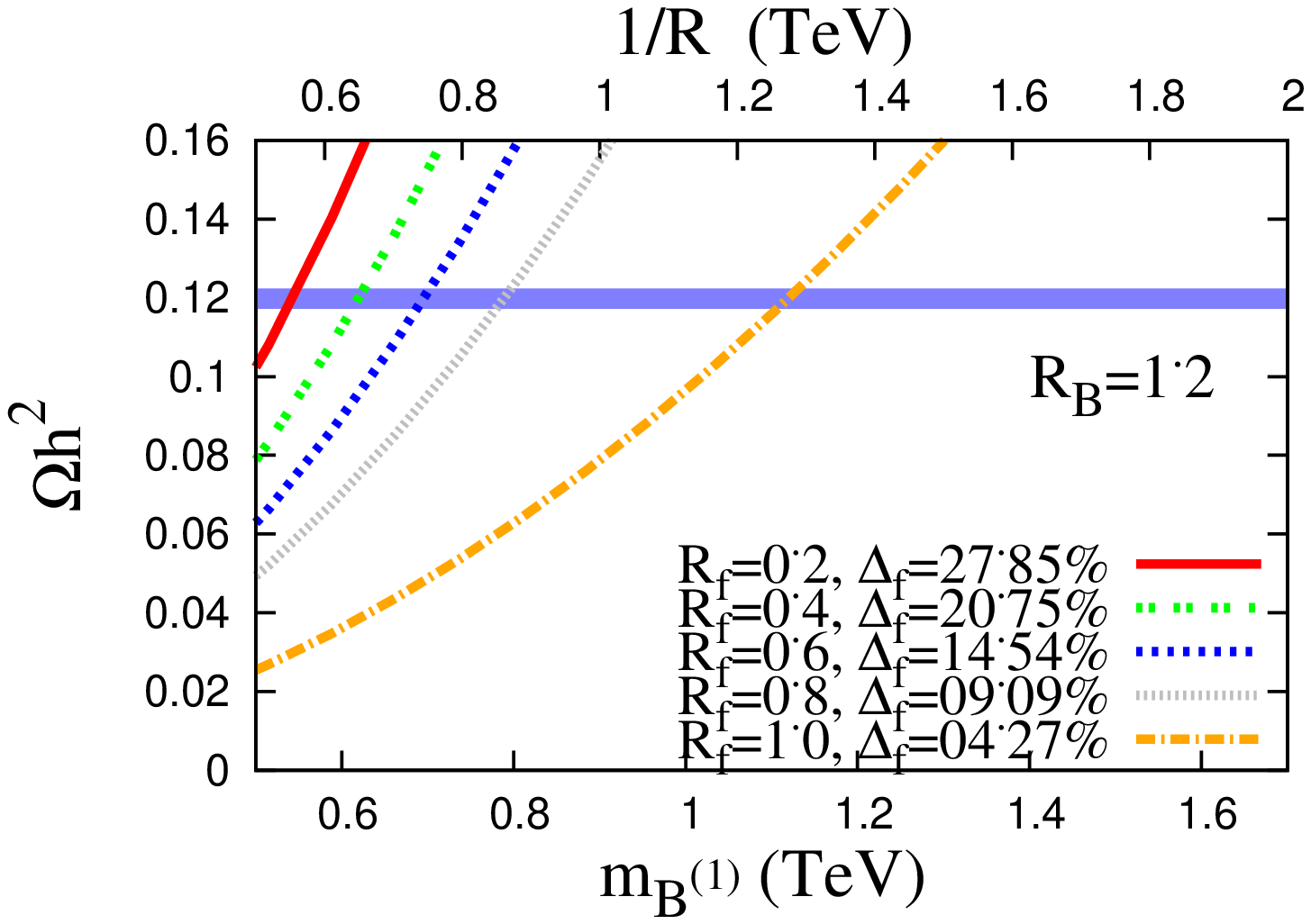}
\caption{Variation of $\Omega h^2$ with relic particle mass.
Curves for different choices of the fermion BLKT parameter  $R_f$
are shown and the corresponding $\Delta_f$ indicated.   The narrow
horizontal blue band corresponds to the 1$\sigma$ allowed range of relic
particle density from Planck data. The allowed $1/R$ can be read
off from the intersections of the curves with the allowed band.
The three panels are for different choices of  $R_B$, the BLKT
parameter for $B$.}

\label{omegahsq} 
\end{center} 
\end{figure}

$B^{(1)}$ is an oft-chosen dark matter candidate in UED. Here for
this choice we use the modifications in the mass spectrum and the
couplings to estimate the relic density in nmUED. In Fig.
\ref{omegahsq} we present the main outcome of this analysis
where in the three panels $R_B$ is chosen differently.  Variation
of $\Omega h^2$ (with the mass of $B^{(1)}$) has been plotted for
different choices of the fermion BLKT parameter, $R_f$. As noted
above, the BLKT parameters for  quarks, gluons and right-handed
fermions is so chosen that the masses of these KK-excitations are
well above that of the $B^{(1)}$.  So, they play no role in the
determination of relic density.  Furthermore, we have checked
that $\Omega h^2$ is nearly insensitive to the choice of $r_h$,
the Higgs BLKT parameter.

It may not be out of place to make a comment about the choice of
parameter values in the plots in Fig. \ref{omegahsq}.  In each 
panel $R_B$ has a fixed value. Thus $m_{B^{(1)}}$ is being varied along
the $x$-axes by changing  $R^{-1}$. To ensure that
$B^{(1)}$ is the LKP, $n=1$ KK-lepton masses should be greater than
$m_{B^{(1)}}$; consequently, $R_f$ (chosen same for all three
left-handed lepton doublets) must be smaller than  $R_B$ in every
panel. In each panel, we also show the values of $\Delta_f$
corresponding to five (equi-spaced) choices of $R_f$.

It would be useful to remark that while calculating
$\sigma_{eff}$ we have taken $l^{(1)}_L$ and $\nu^{(1)}_L$ as mass
degenerate and they are the next heavier in mass than the LKP
assuming all other KK modes being heavy enough so that their
contribution can be neglected.  For a given value of $R^{-1}$, we
find that the cross sections of the reactions listed in Tables
\ref{t:1} - \ref{t:3} are higher compared to the
corresponding rates in UED.  Relatively lower values of the $n=1$
KK-masses and enhanced couplings in nmUED account for the higher
reaction rates for a given value of $R^{-1}$. Consequently, the
numerical value of the relic density in nmUED is always less than in
UED as long as the BLKT parameters are positive (as
used in this analysis).

It is possibly useful to mention here the magnitudes of the
coefficients $I_a$ and $I_b$ in eqn. (\ref{approxomega}) which
reflect the model dynamics.  For the  $R_B$
and $R_f$ considered by us $I_a$ and $I_b$ are in the range
of (1 - 10) $pb$. Both $I_a$ and $I_b$ decrease with increasing
$m_{B^{(1)}}$.

One can see from  Fig. \ref{omegahsq} that $\Omega h^2$
increases with increase of $R^{-1}$ or $R_B$. (Both increases
serve to reduce $m_{B^{(1)}}$ and also enhance the couplings.) 
In contrast, $\Omega h^2$
decreases with increasing $R_f$, the fermion BLKT coefficient.
Unlike in UED, the allowed range of $R^{-1}$ depends on $R_f$ in
this non-minimal version of the model. This is in line with
expectation. As $R_f$ moves away from $R_B$ the splitting,
$\Delta_f$, increases (noted in the figures) and coannihilation
becomes less important.

It is seen from Fig. \ref{omegahsq} that depending on the choice
of $R_B$ and $R_f$ the allowed range of $R^{-1}$ can be as much
as 0.6 TeV to 2 TeV or even more, while remaining consistent with
the observed dark matter limits. It should be borne in mind that
in nmUED the {\em mass} of the LKP is actually a little {\em lower} than
$R^{-1}$ as seen in the lower $x$-axes in the panels.

In passing we remark on the upper bounds on $R^{-1}$
that arise by demanding that the LKP does not overclose the
universe. For this to happen $\Omega h^2$ must be around 0.48.
Some sample results are presented in Table \ref{t:4}. To be
conservative, for every $R_B$ we have chosen the smallest $R_f$
(i.e., the largest $\Delta_f$) in Fig. \ref{omegahsq}. For larger
$R_f$ the upper bound on $R^{-1}$ is increased.
\begin{table}
\vskip -10pt
\begin{center}
\begin{tabular}{|c|c|c|c|c|c|}
\hline
LKP & $R_B$ or $R_W$ & $R_f$ &
$\Delta_f$ & $R^{-1}$  & $m_{B^{(1)}}$ or $m_{W_3^{(1)}}$ \\ 
  & & & & (in TeV)  & (in TeV) \\ \hline
& 0.3 & 0.05 & 7.7\% & 1.9 & 1.7\\ \cline{2-6}
$B^{(1)}$& 0.6 & 0.1 & 15\% & 1.7 & 1.4\\ \cline{2-6}
& 1.2 & 0.2 & 27.8\% & 1.5 & 1.1\\ \hline
$W_3^{(1)}$& 0.6 & - & - & 10.0 & 8.4\\ \hline
\end{tabular}
\caption{Upper bound on $R^{-1}$ from overclosure of the universe
($\Omega h^2$ = 0.48). The mass of the LKP for the limiting
$R^{-1}$ is also presented. For the $W_3^{(1)}$ LKP case only the
process $W_3^{(1)} ~W_3^{(1)} \rightarrow W^{+}~W^{-}$ is
taken into account. Including coannihilation will further enhance
the upper bound in this case.}
\label{t:4}
\end{center}
\end{table}

\begin{figure}[thb] 
\begin{center}{ 
\includegraphics[width=4.8cm,height=4.8cm, angle=270]{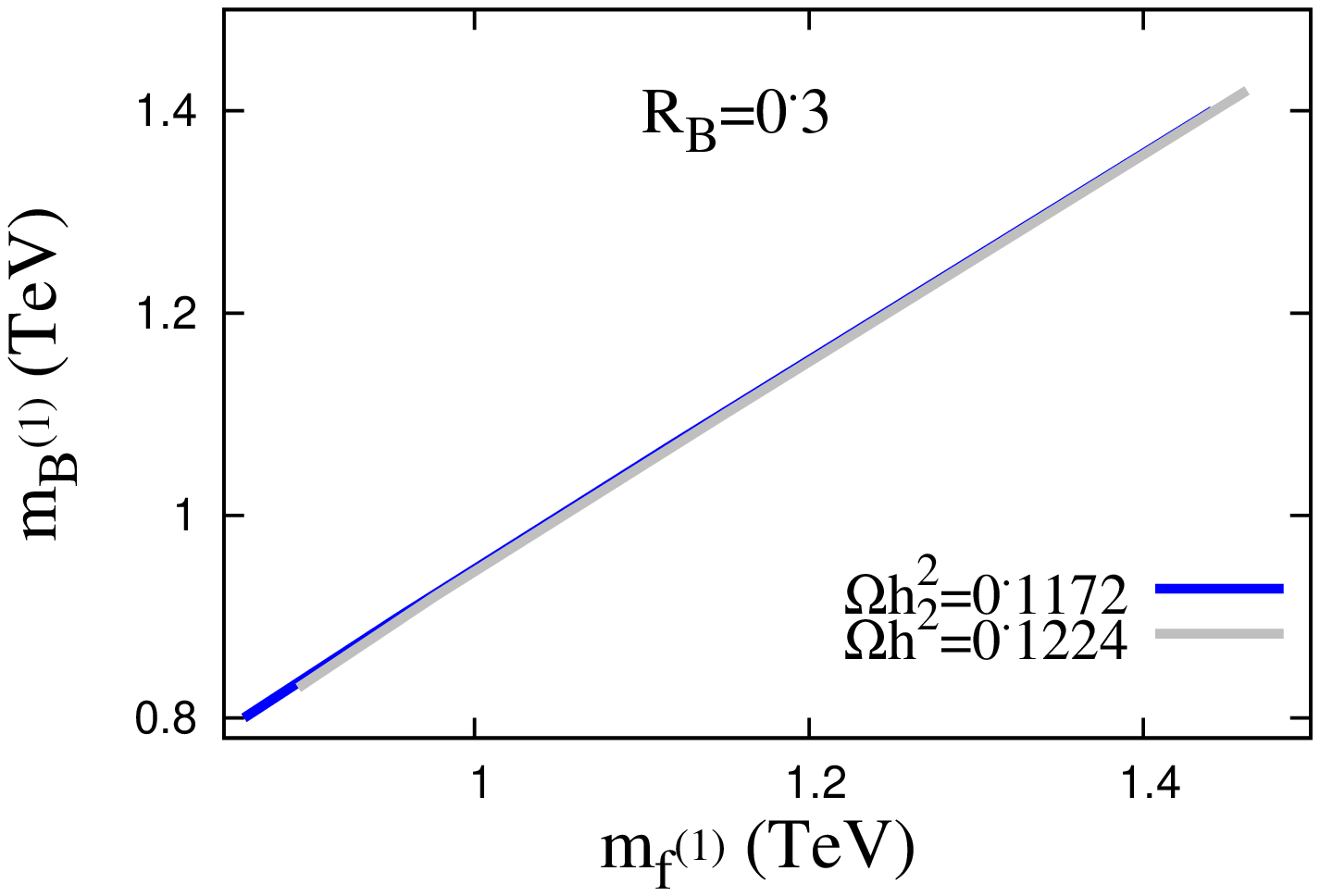}
\includegraphics[width=4.8cm,height=4.8cm,angle=270]{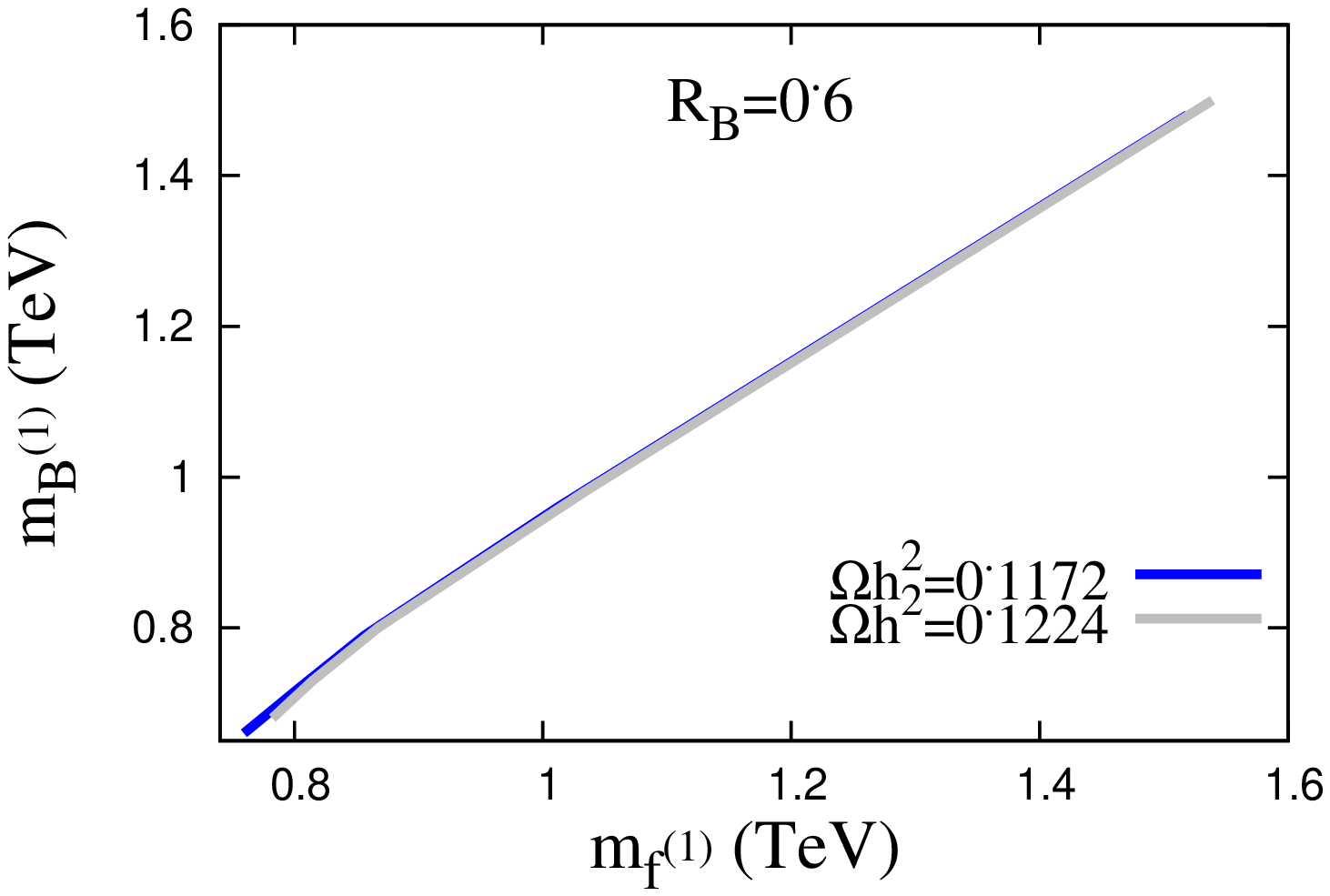} }
\includegraphics[width=4.8cm,height=4.8cm, angle=270]{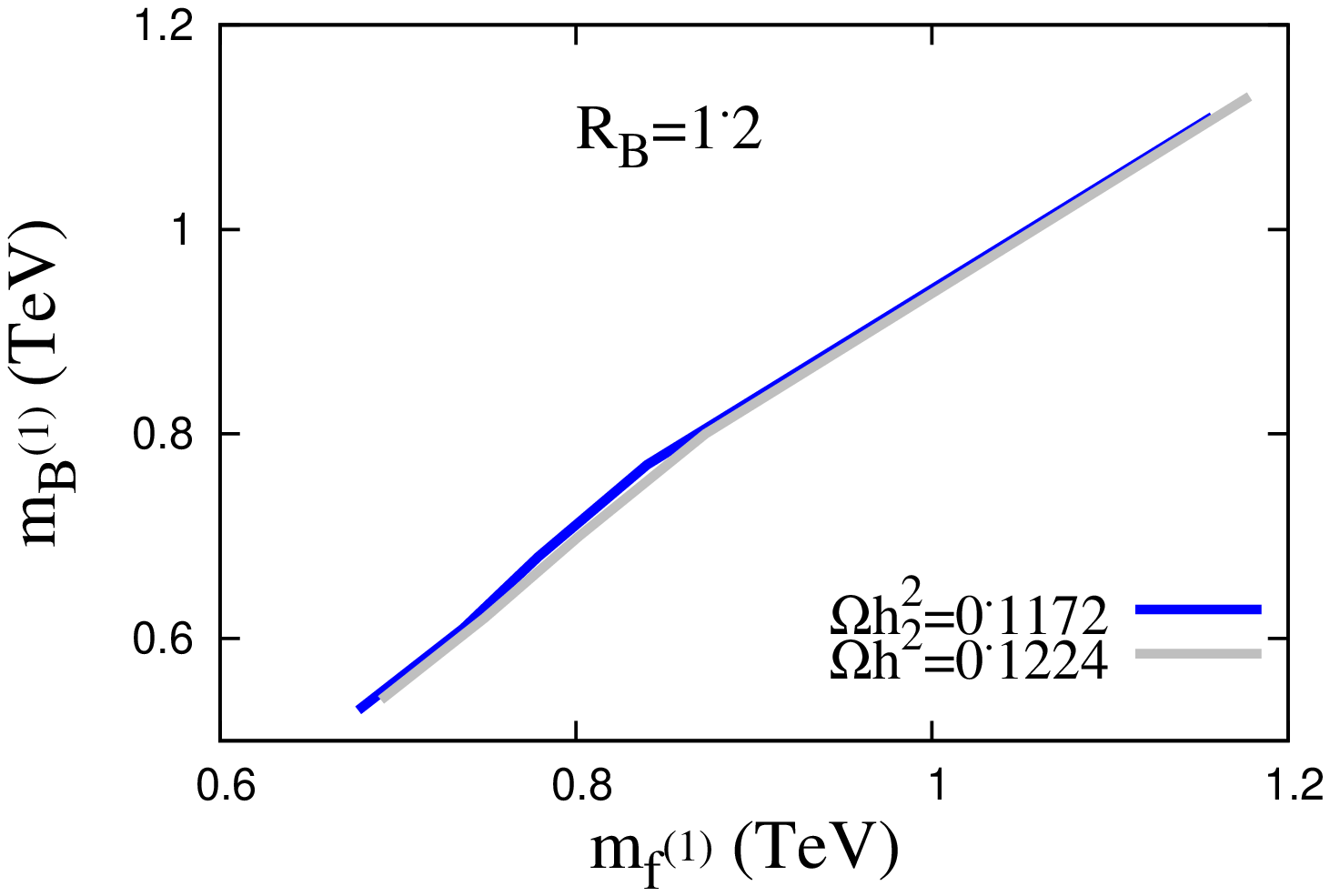}

\caption{Allowed region in the $m_{f^{(1)}} - m_{B^{(1)}}$
plane that satisfies the observed $\Omega h^2$ limits.  The three
panels are for different choices of $R_B$. Only the
narrow strip {\em between} the two curves is allowed from the
relic density constraints.}
\label{allowedrfrb} 
\end{center} 
\end{figure}

The observed $1 \sigma$ limits of $\Omega h^2$, which we have
used, are very restrictive. In Fig. \ref{allowedrfrb} we have
presented the regions in the $m_{B^{(1)}}$-$m_{f^{(1)}}$ plane
which leads to $\Omega h^2$ in the observed range.  While
deriving these limits we assume that $B^{(1)}$ is the {\em only}
relic particle in the model.  Fig. \ref{allowedrfrb} points to
very narrow regions in the $m_{B^{(1)}}$ - $m_{f^{(1)}}$ plane --
{\em between} the two curves in each panel -- allowed by the
data. Thanks to the present precision of cosmological
measurements, when translated to the $m_{B^{(1)}}$-$m_{f^{(1)}}$
plane one has essentially reduced the allowed range to almost a
line. The plots in Fig. \ref{allowedrfrb}
reveal that the allowed range is very close to the $m_{B^{(1)}}
= m_{f^{(1)}}$ line.  (We are not interested in the region above
this line, where $f^{(1)}$ becomes the LKP.)

\subsection{What if $W_3^{(1)}$ is the LKP?}

\begin{figure}[thb] 
\begin{center} 
\includegraphics[width=4.3cm,height=4.8cm,angle=270]{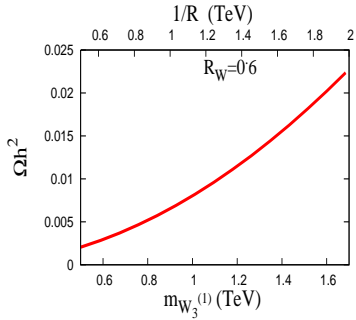}
\caption{Variation of the relic density with $m_{W_3^{(1)}}$ when
$W_3^{(1)}$ is identified as the DM candidate. The current
observed value ($\sim 0.12$) disfavours this alternative.}
\label{Wrelic} 
\end{center} 
\end{figure}

The masses of the $n = 1$ KK-states are determined by their
respective BLKT parameters $r_i$ or equivalently $R_i = r_i/R$.
To avoid a fermion LKP we always choose $R_B, R_W > R_f$. Besides
the just discussed case of  $R_B > R_W$, corresponding to
$B^{(1)}$ being the LKP, one should also consider  $R_W > R_B$
which makes $W_3^{(1)}$ the dark matter candidate\footnote{For a
different discussion of the $W_3^{(1)}$ LKP see, for example,
\cite{ohlsson}.}.  In such an
event we find that the annihilation cross section is large
(notably because of the process $W_3^{(1)} W_3^{(1)}
\rightarrow W^+ W^-$) and therefore the relic density is too
small\footnote{We have not considered coannihilation in this
case. It would further reduce the relic density.}. The relic
density as a function of the $W_3^{(1)}$ mass is shown in Fig.
\ref{Wrelic} for a typical value of $R_W = 0.6$.  Note that the
obtained density is far below the observed \cite{Planck} value
(around 0.12). For $R_W$ = 0.6 the lower (upper) bounds on
$\Omega h^2$ -- which are outside the range of the figure --
correspond to $R^{-1}$ = 4.66 (4.7) TeV with the respective
$W_3^{(1)}$ masses 3.92 (4.04) TeV. This establishes that
$W_3^{(1)}$ would not be an attractive dark matter candidate for
easy detection at the LHC unless the DM has several
components\footnote{The overclosure bound on $R^{-1}$ for $R_W$ =
0.6  is 10 TeV (see Table \ref{t:4}).}.

\section {Direct detection of nmUED relic particle}

Finally, we would like to take a look at the prospects of direct
detection of $B^{(1)}$ at an experiment such as XENON \cite{xenon}.
We will not estimate the actual event rates at such an
experiment; instead, the spin dependent and spin independent
cross sections of $B^{(1)}$ scattering off Xenon nuclei ($A$ = 131,
$Z$ = 54) will be estimated.
Event rates and cross sections of dark matter in the context of
UED have been obtained in ref. \cite{taitservant2, ueddm2}.  As
masses and also the couplings in nmUED show a distinctive
departure from the corresponding quantities in UED, we wish to
relook at the calculation.  

The cross section of scattering of dark matter off nuclei such as
Xenon is ultimately related to its scattering from quarks. As is
customary, we use the non-relativistic limit to separate the
spin-independent (i.e., scalar) and spin-dependent parts of the
cross section.

The scalar part of the scattering cross section of $B^{(1)}$ with a
nucleus of mass $m_N$ with atomic number $Z$ and mass number $A$
at zero momentum transfer is given by
\begin{equation}
\sigma^{scalar}_{0} = \frac{m_N^2}{4 \pi
(m_{B^{(1)}}+m_N)^2}\left(Zf_{p}^{B^{(1)}}+(A-Z)f_{n}^{B^{(1)}}\right)^{2},
\label{sigma0}
\end{equation}
where
\begin{equation}
f_{p,n}^{B^{(1)}} = m_{p,n}\sum_{q}
\frac{\gamma_{q}+\beta_{q}}{m_{q}} f_{T_{q}}^{(p,n)} \ .
\end{equation}
In our numerical evaluations  we have used $f_{T_{q}}^{(p,n)} =
\langle q\bar{q} \rangle_{p,n} (m_q/m_{p,n})$, which relate the
quark-level cross sections to that for the nucleons, as given in
\cite{jungman}.  The physics of the specific dark matter
candidate is captured in the quantities $\gamma_{q}$ and $\beta_{q}$
which here stand for the interaction of $B^{(1)}$ with quarks
mediated via the SM Higgs exchange and $n=1$ KK-fermion exchange
respectively. We find
\begin{equation}
\gamma_{q}  = -\frac{g^{\prime 2}_{1}}{2}\frac{m_{q}}{m_h^2}\ ,
~~{\rm and}~~
\beta_{q}  =
-m_{q}\frac{\tilde{g}_{1}^{2}\left(Y_{qL}^2+Y_{qR}^2\right)}
{\left(m_{B^{(1)}}^{2}-m_{q^{(1)}}^{2}\right)^2}\left(m_{B^{(1)}}^{2}
+m_{f^{(1)}}^{2}\right)\ .  
\end{equation}
where 
\begin{equation}
g^{\prime 2}_1 =  g_1^2   ~\pi R
\left(1+\frac{R_B}{\pi}\right) ~\frac{1}{\sqrt{1+\frac{R_h}{\pi}}}
~I_{B^{(1)}B^{(1)}h^{(0)}}
\end{equation}
where $I_{B^{(1)}B^{(1)}h^{(0)}}$ is given in eqn. \ref{A4}. 
\begin{equation}
\tilde{g}_1 = g_{1}\sqrt{\pi R \left(1+\frac{R_B}{\pi}\right)}
~I_{B^{(1)}f^{(1)}f^{(0)}} \ ,
\end{equation}
with $I_{B^{(1)}f^{(1)}f^{(0)}}$ given in  eqn. \ref{A1}.
Numerically $\gamma_q$ is almost insensitive to $m_{B^{(1)}}$ and
about two orders of magnitude larger than $\beta_q$.

The corresponding spin-dependent cross section is given by:
\begin{equation}
\sigma^{spin}_{0} = \frac{2}{3 \pi}
(\mu^2 \tilde{g}_1^4) \left(\frac{a_p \langle S_p \rangle + a_n \langle S_n
\rangle}{m_{B^{(1)}}^2 - m_{q^{(1)}}^2}\right)^2 \frac{(J+1)}{J}
\label{sigma0sp}
\end{equation}
where $\mu$ is the reduced mass of the target nucleus and
the the other nuclear parameters can be found in \cite{jungman}.
In particular, 
\begin{equation}
a_p  =  \frac{17}{36}\Delta u + \frac{5}{36}(\Delta d+\Delta s)
~~{\rm and }~~ a_n =  \frac{17}{36}\Delta d + \frac{5}{36}(\Delta
u+\Delta s) 
\end{equation}
Following ref. \cite{taitservant2}, in our analysis  we have used
the central values of $\Delta u = 0.78 \pm 0.02$, $\Delta d =
-0.48 \pm 0.02$ and $\Delta s = -0.15 \pm 0.02$.

Experimental results are often presented in terms of {\em
effective} dark matter - nucleon scattering cross sections given by:
\begin{eqnarray}
\sigma_{p,n}^{scalar} &=& \sigma_0 \frac{m_{p,n}^2}{\mu^2}\frac{1}{A^2}  .
\ , \nonumber
\\
\sigma_{p,n}^{spin} &=& \frac{\tilde{g}_1^4}{2
\pi}\frac{\mu_{p,n}^2 a_{p,n}^2}{(m_{B^{(1)}}^2-m_{f^{(1)}}^2)^2}
\label{DDxsections}
\end{eqnarray}
Here, $\mu_{p,n}$ is the reduced mass of the
WIMP-nucleon  system.

\begin{figure}[thb] 
\begin{center} 
{\includegraphics[width=4.8cm,height=4.8cm,angle=0]{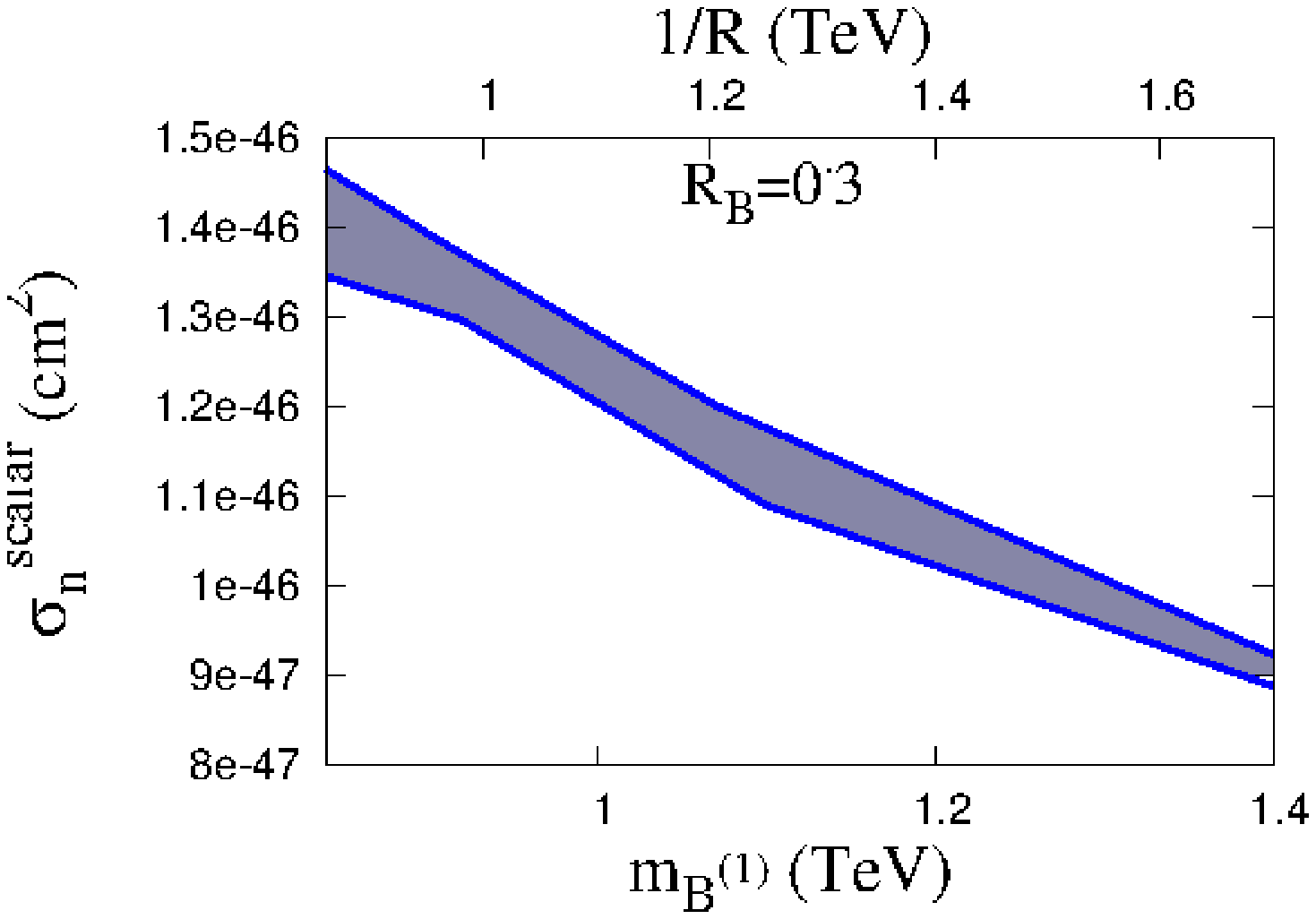} 
\includegraphics[width=4.8cm,height=4.8cm, angle=0]{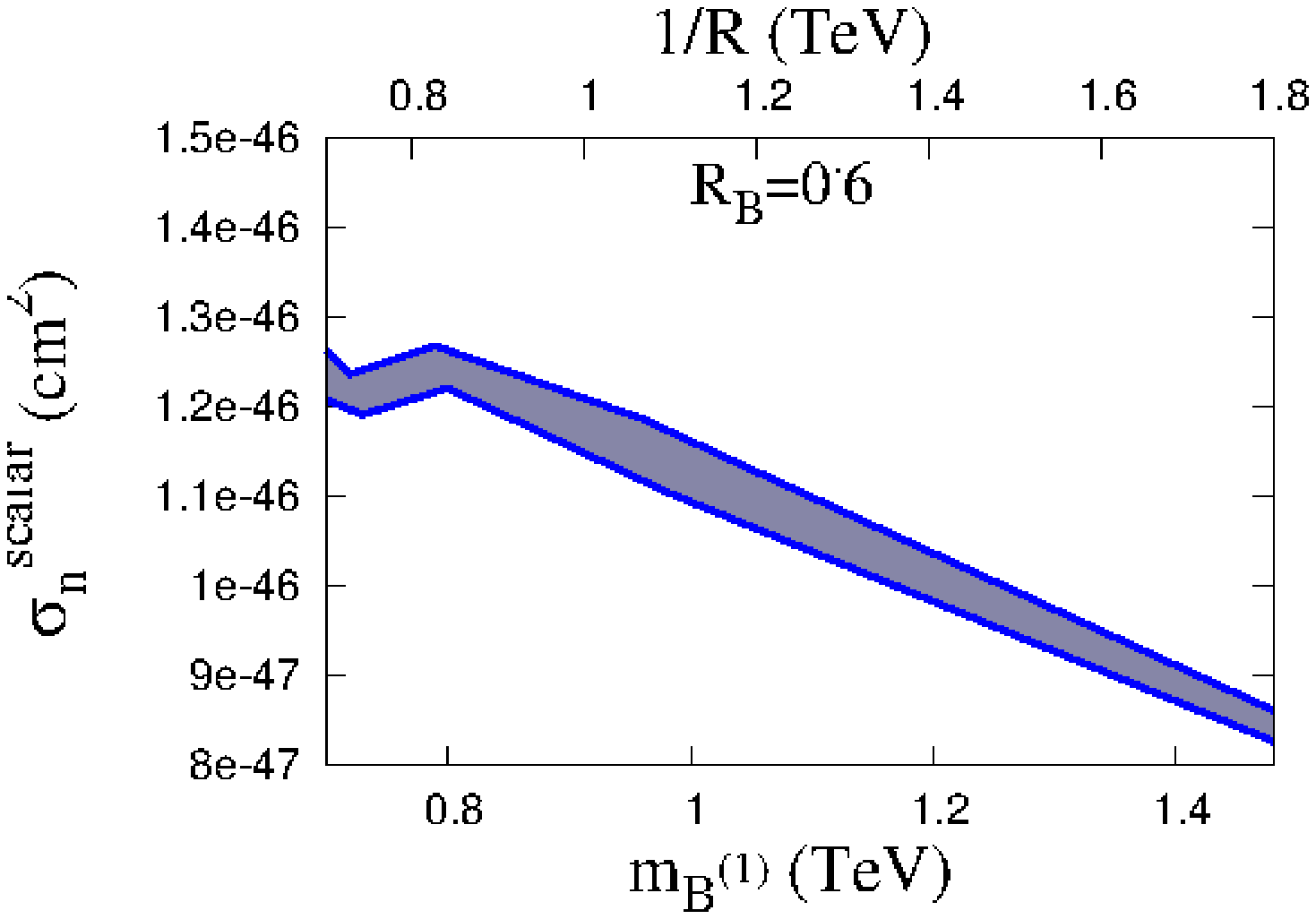}}
\includegraphics[width=4.8cm,height=4.8cm, angle=0]{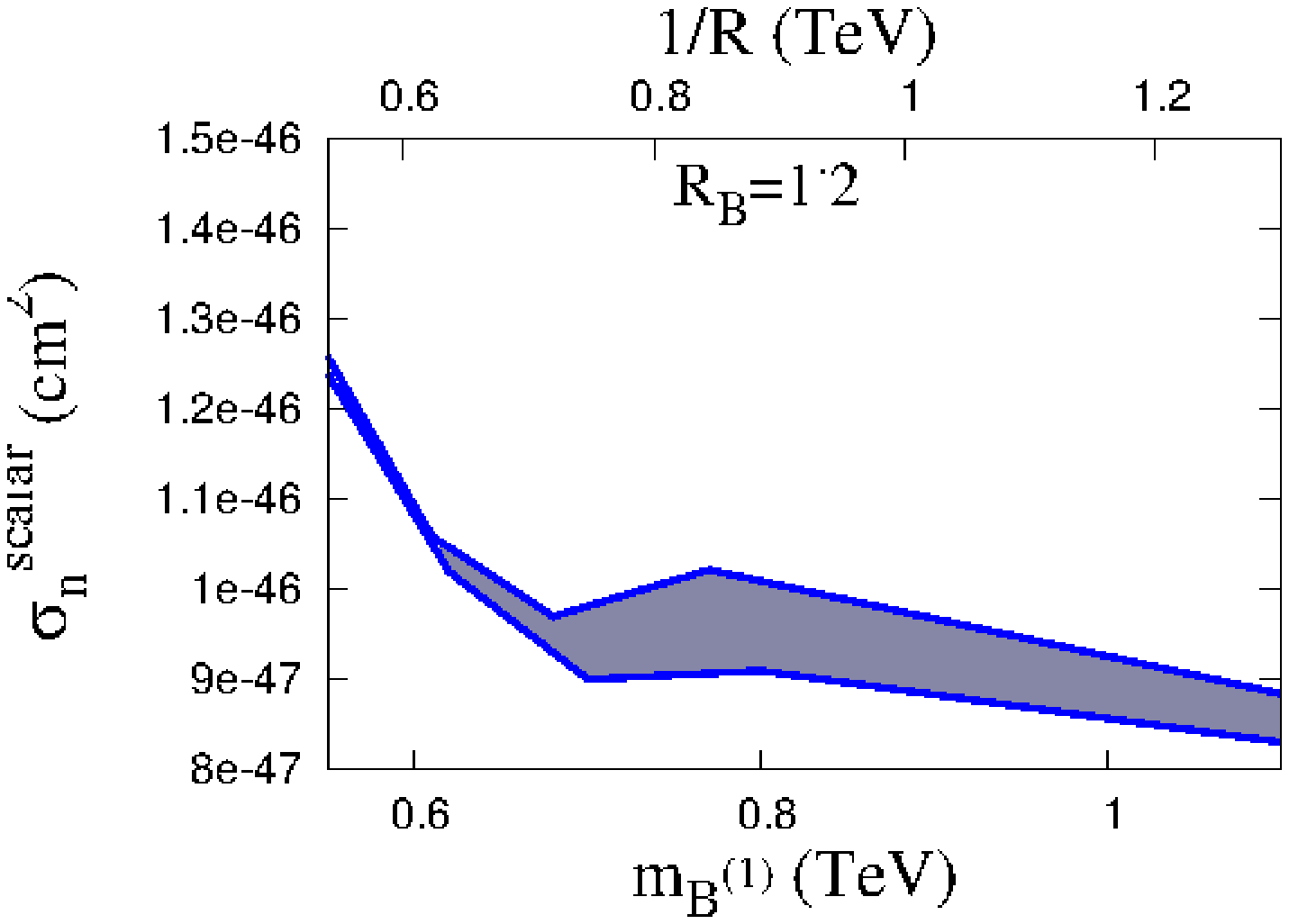}
\caption{Variation of the spin independent WIMP nucleon cross section
with relic particle mass for Xenon. The three panels are for
three values of $R_B$. The shaded region is obtained by using the BLKT
parameters consistent with the observed relic density.  }
\label{DD_scalar} 
\end{center} 
\end{figure} 

\begin{figure}[thb] 
\begin{center} 
{\includegraphics[width=5.2cm,height=4.8cm,angle=0]{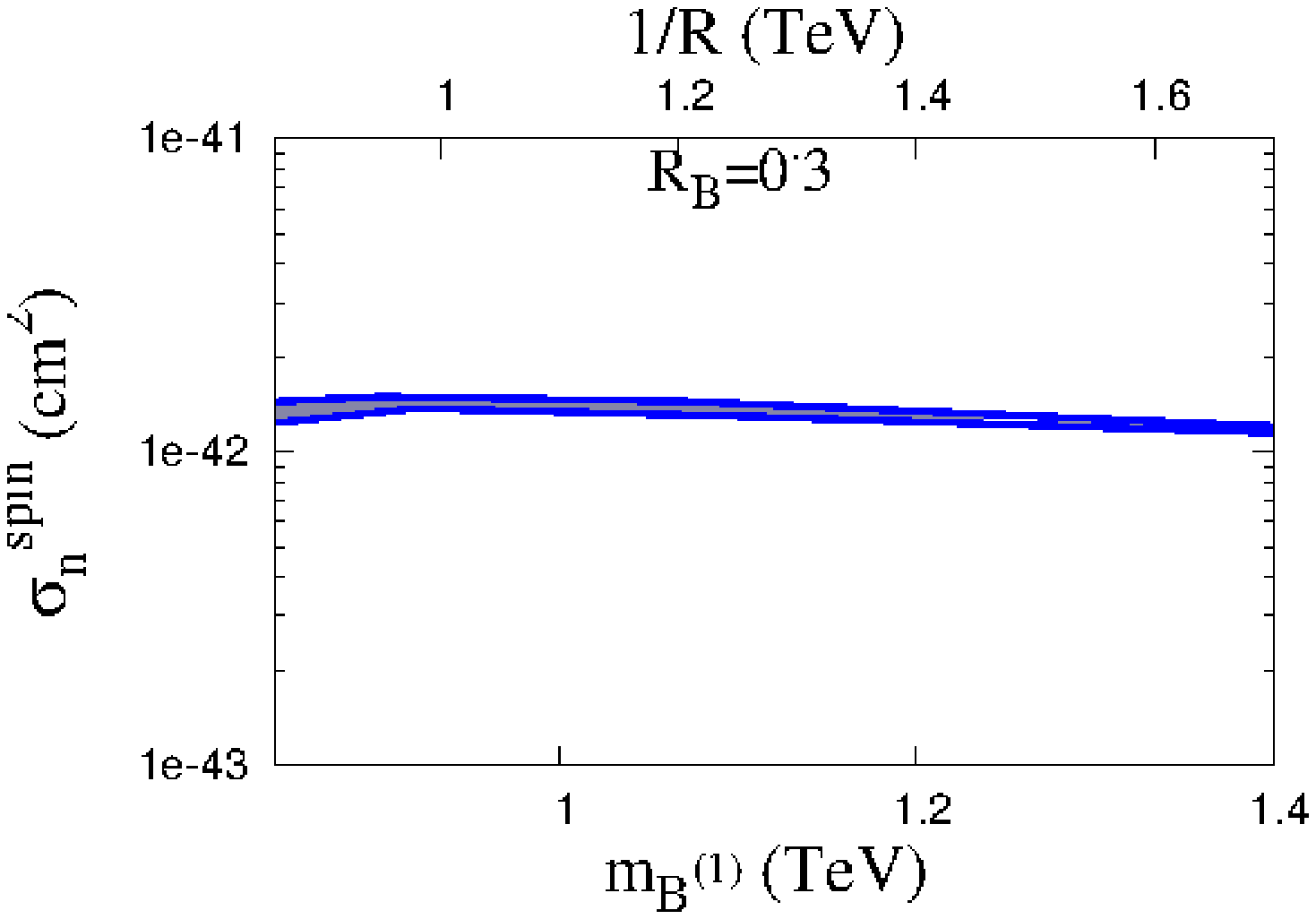} 
\includegraphics[width=5.2cm,height=4.8cm, angle=0]{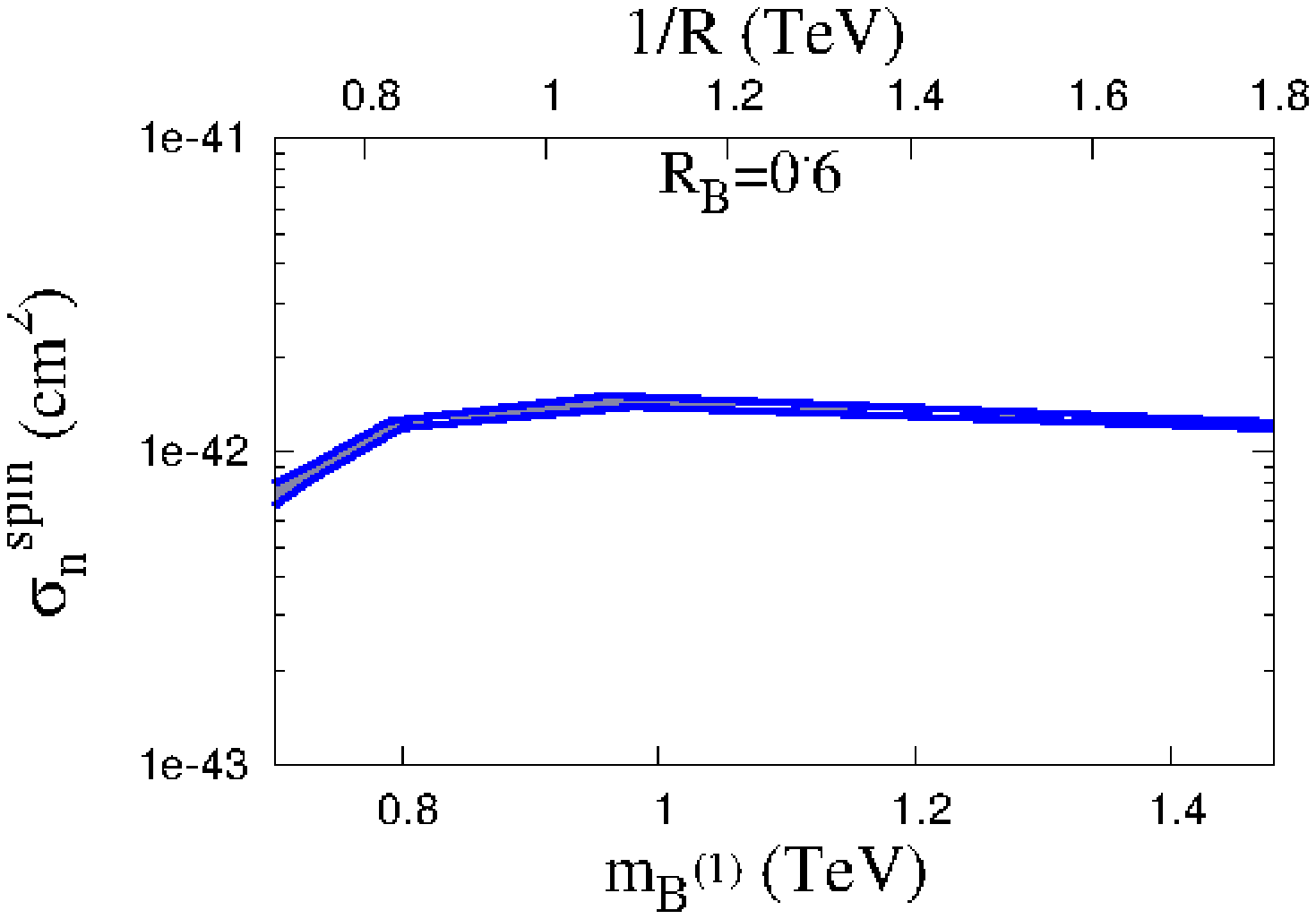}}
\includegraphics[width=5.2cm,height=4.8cm, angle=0]{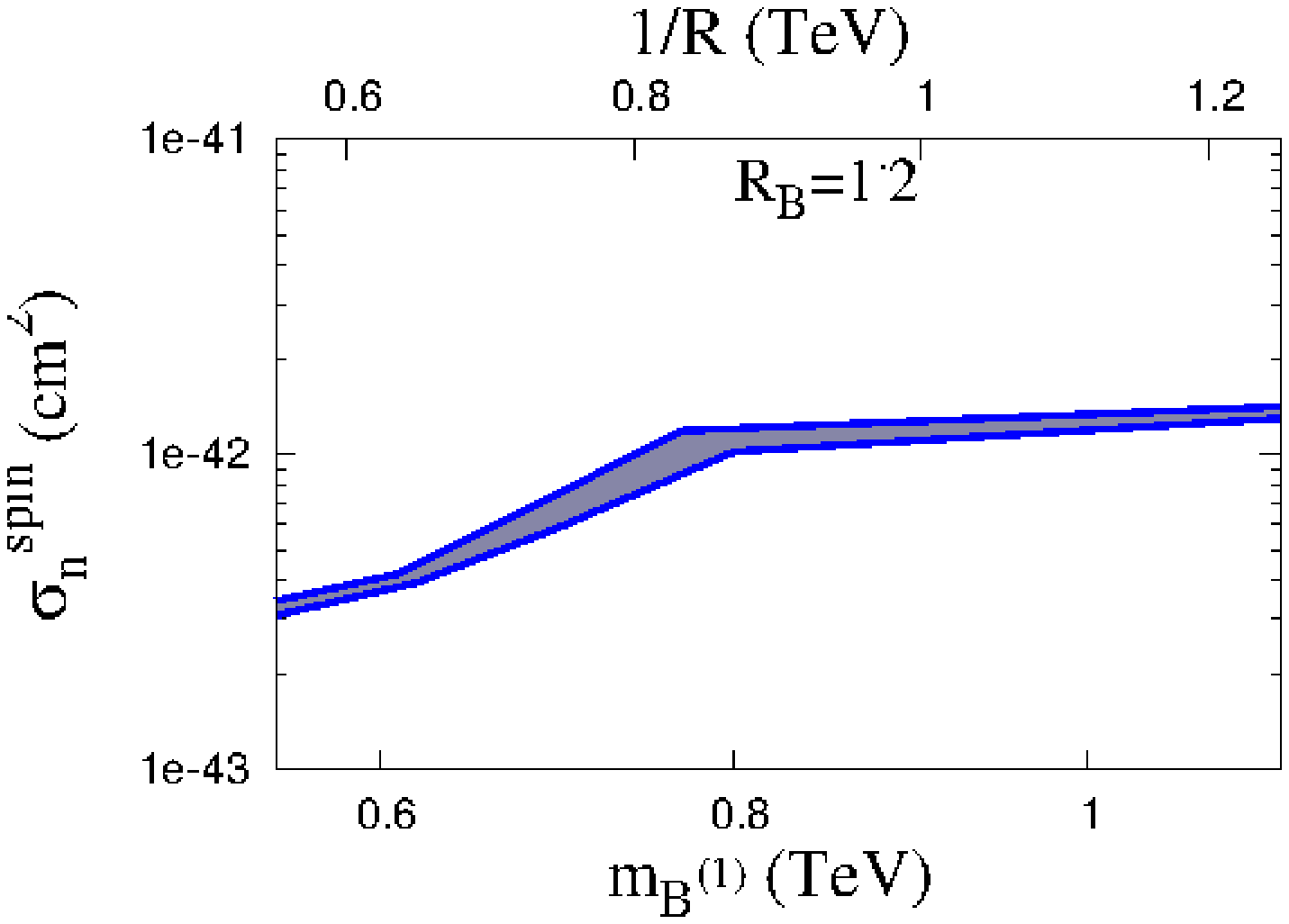}
\caption{Variation of the spin dependent WIMP nucleon cross section
with relic particle mass for Xenon. The three panels are for
three values of $R_B$. The shaded region is obtained by using the BLKT
parameters consistent with the observed relic density. }
\label{DD_spin} 
\end{center} 
\end{figure} 

We can now turn to the numerical results for the spin-dependent
and scalar WIMP-nucleon cross section for Xenon.  The  cross
sections will be presented as a function of the LKP ($B^{(1)}$)
mass which is fixed by $R^{-1}$ and the
BLKT parameter $R_B$.  We will restrict $R_f$ the lepton BLKT
strength to only those values which satisfy the relic density
$\Omega h^2$ requirement once $R_B$ is fixed; i.e., the range
shown in Fig. \ref{allowedrfrb}.  To appreciate this choice of
input parameters we refer to Fig. \ref{omegahsq}.  For a fixed
value of $R_B$ (i.e., any one panel) and $R_f$ 
(one of the curves in the panel) one has two values of $R^{-1}$
resulting from the 1$\sigma$ upper and lower allowed values of
$\Omega h^2$. For each such $R_B, R_f$ pair the  (two) $R^{-1}$
values have been used to present the direct detection cross
section bands in Figs. \ref{DD_scalar} and \ref{DD_spin}.  Thus
as we go along left to right along $x$-axis of these plots, $R_f$
increases  along with
$m_{B^{(1)}}$.

Scattering rates of the the LKP WIMP off nucleons are
presented in Fig. \ref{DD_scalar} (scalar cross section,
$\sigma_n ^{scalar}$) and in Fig. \ref{DD_spin} (spin dependent
cross section, $\sigma_n ^{spin}$).  Over the range of input
parameters we have considered, $\sigma_n^{scalar}$ decreases by
two orders of magnitute from 10$^{-46}$ cm$^2$ to 10$^{-48}$
cm$^2$ while $\sigma_n^{spin}$ is around 10$^{-42}$ cm$^2$
and is not very sensitive to the change of $R^{-1}$ and $R_f$.
It would be relevant here to mention that these values of the
WIMP scattering cross sections are well below the sensitivity of
the XENON experiment\footnote{For the range of WIMP masses in
Fig. \ref{DD_scalar}, sensitivity of the XENON experiment
\cite{xenon}  for the scalar cross section is above $10^{-45}$
cm$^2$ \cite{xenon}.}.  Thus the entire range of parameters used
for the direct detection rates is allowed by the XENON
experiment.

The plots in Figs. \ref{DD_scalar} and \ref{DD_spin}  for the
dark matter cross sections appear somewhat different from those
given for UED in \cite{taitservant2} and \cite{kongmatchev2}. The
reason for this is that unlike in those analyses here the fermion
KK-excitation masses and the $B^{(1)}$ mass are strongly
correlated (see Fig. \ref{allowedrfrb}) by virtue of the
requirement that the observed DM relic density be reproduced.

The nature of variation of the DM scattering  cross sections as
presented in Figs.  \ref{DD_scalar} and \ref{DD_spin} can by seen
to follow from eqn. (\ref{DDxsections}). BLKT parameter
dependence creeps into $\sigma_n^{spin}$ via the $\left(
m_{f^{(1)}} ^2 - m_{B^{(1)}}^2\right) ^2$ factor in the
denominator along with $\tilde{g}_1$. $\tilde{g}_1$ is barely
senisitive to $R^{-1}$ or $R_f$.  And so is the former: as we go
from left to right along the $x$-axis, $\left( m_{f^{(1)}} -
m_{B^{(1)}} \right)$ decreases while $\left( m_{f^{(1)}} +
m_{B^{(1)}} \right)$  increases, making the cross section
almost independent of $m_{B^{(1)}}$. On the other hand, the
scalar cross section has a more complicated dependence on the
BLKT parameters.  While $\gamma_q \over m_q$ is almost
independent of BLKT parameters, $\beta_q \over m_q$ increases
with $m_{B^{(1)}}$ and $R_f$. However, the $\gamma_q$
contribution dominates  and thus the combination $(-\gamma_q -
\beta_q) \over m_q$ changes slowly with $m_{B^{(1)}}$ and $R_f$.
An overall factor of $m_{B^{(1)}}^2$ in the denomintor of
$\sigma_0$ (see eqn. (\ref{sigma0})), accounts for the rapid
decrease of the scalar cross section, which falls monotonically
in Fig. \ref{DD_scalar}.

\section{Summary and Conclusions}

Universal Extra Dimension models have emerged as an attarctive
option for Beyond the Standard Model physics. In this model all
SM particles are complemented with KK-excitations which are
equispaced in mass. The interaction strengths of these states are
determined entirely by the SM. Many aspects of the
model ranging from constraints from precision measurements to
collider searches have been examined in the literature. Signals for UED are
being actively looked for at the LHC.

One of the less attractive predictions of UED is the mass
degeneracy of KK-excitations of all SM particles at any fixed
level. A remedy for this is not unknown. It has been shown
\cite{cms1} that five-dimensional radiative corrections split the
degeneracy in a definite way determined by the SM charges of the
zero-mode states. The corrections are encoded as contributions
to the four-dimensional lagrangians located at the the two fixed
points of the orbifold.  In this version of the model, known as
minimal UED,  the practice has been to assume that the
couplings of the KK excitations continue to be as for the SM
particles.
 
Our work is on a further generalisation of this model where the
extra four-dimensional kinetic terms located at the two fixed
points are of a strength whch is a free parameter and varies from
particle to particle.  To ensure the conservation of a $\mathbf{Z}_2$
parity the strengths are taken equal at the two fixed points. This
ensures the stability of the LKP. The
BLKT parameters determine the wave-functions of the
KK-excitations in the fifth dimension, $y$, as well as their masses. 
Further, the non-trivial $y$-dependence of the wave-functions
affects the couplings of the KK-excitations; these are also controlled
by the BLKT parameters. We allow different BLKT strengths for the
various SM particles and ensure that $B^{(1)}$ is the LKP. We
also examine the alternative of a $W_3^{(1)}$ LKP but find that
the relic density is too small for a WIMP mass $\sim$ 1 TeV. We
conclude that $W_3^{(1)}$ cannot serve the role of a single
component dark matter when its mass is within the LHC range. We
make a note of the bounds on the LKP dark matter particle mass
which follow from the overclosure of the universe.

In this work we consider dark matter in this nmUED scenario
retaining the impact of BLKT parameters on the masses {\em
and} the couplings. We show that the  range of relic densities
preferred by the Planck data places stringent restrictions on the
BLKT strengths of the gauge bosons and fermions and these get
correlated. We find that in this process the allowed range  of the
compactification scale $R^{-1}$ is much relaxed from its narrow
UED  prediciton of 0.5-0.6 TeV .  

We discuss the prospects of direct detection of the nmUED dark
matter candidate keeping the relic density constraints in mind.
As an example, we evaluate the spin-dependent and
spin-independent scattering cross section of 
dark matter off Xenon nuclei. Our calculations indicate
that the signal is well below the existing limits set by
XENON100 for spin-independent  scattering.

{\bf Acknowledgements:} The authors are grateful to Tirtha Sankar
Ray for collaborating in the early stages of this work. U.K.D. and
A.S. thank Kirtiman Ghosh for valuable discussions. A.D. acknowledges partial support from the
DRS project sanctioned to the Department of Physics, University
of Calcutta by the University Grants Commission. U.K.D. is supported
by  funding from the Department of Atomic Energy, Government of
India for the Regional Centre for Accelerator-based Particle
Physics, Harish-Chandra Research Institute (HRI).  A.R. is
partially funded by  the Department of Science and Technology
Grant No. SR/S2/JCB-14/2009.   A.S. thanks the University
Grants Commission for support. 
 
{\Large{\bf Appendix A: $n = 1$ KK-excitation Feynman rules}}
\setcounter{equation}{0}  
\setcounter{section}{1}  

\renewcommand{\thesection}{\Alph{section}}
\renewcommand{\theequation}{\thesection-\arabic{equation}}  

In this Appendix we note the Feynman rules for the
$n = 1$ Kaluza-Klein excitations. Each of these vertices involves
a nontrivial coupling determined by the five-dimensional
wave-functions of the KK-excitations. These couplings are 
listed separately below. Besides these Feynman rules and couplings,
only the SM rules are required.

\subsection{Feynman Rules:}

\subsubsection{\bf The $B^{(1)} f^{(1)} f^{(0)}$ vertex }\label{b1f1f0}
\vspace{-0.5 cm}
\fcolorbox{white}{white}
{  \begin{picture}(244,166) (15,-91)
    \SetWidth{1.0}
    \SetColor{Black}
    \Photon(20,-10)(70,-10){4}{8}
    \Line[arrow,arrowpos=0.5,arrowlength=5,arrowwidth=2,arrowinset=0.2](70,-10)(120,30)
    \Line[arrow,arrowpos=0.5,arrowlength=5,arrowwidth=2,arrowinset=0.2](120,-50)(70,-10)
    \Text(130,-15)[lb]{\Black{$\equiv-i\left(g_{1}\sqrt{\pi R
\left(1+\frac{R_B}{\pi}\right)}\times
I_{B^{(1)}f^{(1)}f^{(0)}}\right)\gamma^{\mu}(P_LY_L+P_RY_R)$}}
    \Text(39,0)[lb]{\Black{$B^{(1)}$}}
    \Text(80,-50)[lb]{\Black{$\bar{f}^{(1)}$}}
    \Text(80,10)[lb]{\Black{$f^{(0)}$}}
  \end{picture}
}
\subsubsection{\bf The $B^{(1)} h^{(1)} h^{(0)}$ vertex }\label{b1h1h0}
\vspace{-0.5 cm}
{  \begin{picture}(244,166) (15,-91)
    \SetWidth{1.0}
    \SetColor{Black}
    \Photon(20,-10)(70,-10){4}{8}
    \Line[dash,dashsize=5,arrow,arrowpos=0.5,arrowlength=5,arrowwidth=2,arrowinset=0.2](70,-10)(120,30)
    \Line[dash,dashsize=5,arrow,arrowpos=0.5,arrowlength=5,arrowwidth=2,arrowinset=0.2](120,-50)(70,-10)
    \Text(130,-15)[lb]{\Black{$\equiv-i\left(g_{1}\sqrt{\pi R \left(1+\frac{R_B}{\pi}\right)}\times I_{B^{(1)}h^{+(1)}h^{-(0)}}\right)Y_H(p_{\mu}^{-}-p_{\mu}^{+})$}}
    \Text(39,0)[lb]{\Black{$B^{(1)}$}}
    \Text(110,-35)[lb]{\Black{$p^{+}$}}
    \Text(80,-50)[lb]{\Black{$h^{+(1)}$}}
    \Text(110,10)[lb]{\Black{$p^{-}$}}
    \Text(77,10)[lb]{\Black{$h^{-(0)}$}}
  \end{picture}}
\subsubsection{\bf The $B^{(1)} B^{(1)} h^{(0)} h^{(0)}$ vertex }
\label{b1b1h0h0}
\vspace{0.5 cm}
{  \begin{picture}(244,166) (15,-91)
    \SetWidth{1.0}
    \SetColor{Black}
    \Photon(20,80)(70,34){4}{8}
    \Photon(70,34)(20,-20){4}{8}
    \Line[dash,dashsize=5,arrow,arrowpos=0.5,arrowlength=5,arrowwidth=2,arrowinset=0.2](70,34)(120,80)
    \Text(40,70)[lb]{\Black{$B^{(1)}$}}
    \Text(40,-10)[lb]{\Black{$B^{(1)}$}}
    \Line[dash,dashsize=5,arrow,arrowpos=0.5,arrowlength=5,arrowwidth=2,arrowinset=0.2](120,-20)(70,34)
    \Text(80,70)[lb]{\Black{$h^{+(0)}$}}
    \Text(72,-10)[lb]{\Black{$h^{-(0)}$}}
    \Text(130,25)[lb]{\Black{$\equiv-2i \left(g^2_1 \pi R \left(1+\frac{R_B}{\pi}\right) \times I_{B^{(1)}B^{(1)}h^{+(0)}h^{-(0)}}\right)\eta_{\mu\nu}Y^2_H$}}
  \end{picture}
}
\vspace{-0.5 cm}
\subsubsection{\bf The $B^{(1)} B^{(1)} h^{(0)}$ vertex }\label{b1b1h0}
\vspace{-0.5 cm}
{
  \begin{picture}(232,127) (34,-168)
    \SetWidth{1.0}
    \SetColor{Black}
    \Photon(35,-55)(85,-104){4}{8}
    \Photon(85,-104)(42,-167){4}{8}
    \Line[dash,dashsize=5,arrow,arrowpos=0.5,arrowlength=5,arrowwidth=2,arrowinset=0.2](85,-104)(145,-104)
    \Text(65,-73)[lb]{\Black{$B^{(1)}$}}
    \Text(65,-150)[lb]{\Black{$B^{(1)}$}}
    \Text(110,-97)[lb]{\Black{$h^{(0)}$}}
    \Text(150,-120)[lb]{\Black{$\equiv -2i \left(g_{1}^{2}\pi R
\left(1+\frac{R_B}{\pi}\right)\times
I_{B^{(1)}B^{(1)}h^{(0)}}\right)\times
\frac{v}{\sqrt{1+\frac{R_h}{\pi}}} Y_H^2$}} \end{picture} } 
\subsubsection{\bf The $W^{(1)} f^{(1)} f^{(0)}$ vertices }\label{w1f1f0}
\vspace{-0.5 cm}
{  \begin{picture}(244,166) (15,-91)
    \SetWidth{1.0}
    \SetColor{Black}
    \Photon(20,-10)(70,-10){4}{8}
    \Line[arrow,arrowpos=0.5,arrowlength=5,arrowwidth=2,arrowinset=0.2](70,-10)(120,30)
    \Line[arrow,arrowpos=0.5,arrowlength=5,arrowwidth=2,arrowinset=0.2](120,-50)(70,-10)
    \Text(130,-15)[lb]{\Black{$\equiv-i\left(g_{2}\sqrt{\pi R \left(1+\frac{R_W}{\pi}\right)}\times I_{W^{(1)}f^{(1)}f^{(0)}}\right)\gamma^{\mu} T_{3} P_L$}}
    \Text(39,0)[lb]{\Black{$W_{3}^{(1)}$}}
    \Text(80,-50)[lb]{\Black{$\bar{f}^{(1)}$}}
    \Text(80,10)[lb]{\Black{$f^{(0)}$}}
  \end{picture}
}
\vspace{-2.5 cm}
{  \begin{picture}(244,166) (15,-91)
    \SetWidth{1.0}
    \SetColor{Black}
    \Photon(20,-10)(70,-10){4}{8}
    \Line[arrow,arrowpos=0.5,arrowlength=5,arrowwidth=2,arrowinset=0.2](70,-10)(120,30)
    \Line[arrow,arrowpos=0.5,arrowlength=5,arrowwidth=2,arrowinset=0.2](120,-50)(70,-10)
\Text(130,-15)[lb]{\Black{$\equiv-i\left(g_2\sqrt{\pi R \left(1+\frac{R_W}{\pi}\right)}\times I_{W^{(1)}f^{(1)}f^{(0)}}\right) \frac{\gamma^{\mu}}{\sqrt{2}} P_L$}}
    \Text(39,0)[lb]{\Black{$W^{+ (1)}$}}
    \Text(80,-50)[lb]{\Black{$\bar{f}^{(1)}$}}
    \Text(80,10)[lb]{\Black{$f^{(0)}$}}
  \end{picture}
}
\vspace {-2.0cm}
\paragraph*{}

\subsection{Couplings:} 

Here we list the couplings which appear in the Feynman rules
given above. 
\begin{equation}
I_{B^{(1)}f^{(1)}f^{(0)}} = \int_{0}^{\pi R} \left[1+r_f
\{\delta(y)+\delta(y-\pi R)\}\right]a_B^{(1)}f^{(1)}f^{(0)} dy.
\label{A1}
\end{equation}
\begin{equation}
I_{B^{(1)}h^{(1)}h^{(0)}} = \int_{0}^{\pi R} \left[1+r_h
\{\delta(y)+\delta(y-\pi R)\}\right] a_B^{(1)}h^{(1)}h^{(0)} dy .
\label{A2}
\end{equation}
\begin{equation}
I_{B^{(1)2}h^{(0) 2}} = \int_{0}^{\pi R} \left[1+r_h
\{\delta(y)+\delta(y-\pi R)\}\right] a_B^{(1) 2} h^{(0)2} dy .
\label{A3}
\end{equation}
\begin{equation}
I_{B^{(1)}B^{(1)}h^{(0)}} = \int_{0}^{\pi R} \left[1+r_h
\{\delta(y)+\delta(y-\pi R)\}\right]a_B^{(1)}a_B^{(1)}h^{(0)} dy .
\label{A4}
\end{equation}
\begin{equation}
I_{W^{(1)}f^{(1)}f^{(0)}} = \int_{0}^{\pi R} \left[1+r_f
\{\delta(y)+\delta(y-\pi R)\}\right]a_W^{(1)}f^{(1)}f^{(0)} dy.
\label{A5}
\end{equation}
where $a^{(n)}_B(y), f^{(n)}(y)$ are the the wave-functions for the gauge
boson and fermion fields introduced earlier, and $h^{(n)}(y)$ is
the same for the Higgs field.

\end{document}